\documentclass[useAMS,usenatbib]{mn2e}
\usepackage{aas_macros}
\usepackage{times}
\usepackage{amssymb}
\usepackage{graphicx}
\usepackage{rotating}
\usepackage{multirow}
\begin{document}

\title[A submm survey of LABs in the SA\,22 protocluster]
{A submm survey of Ly$\alpha$ haloes in the SA\,22 protocluster at $z=3.1$}
\author[J. E. Geach et al]
{\parbox[h]{\textwidth}{
J.\,E.\,Geach$^1$\thanks{E-mail: j.e.geach@durham.ac.uk}, Y.\,Matsuda$^2$, Ian\,Smail$^1$, S.\,C.\,Chapman$^3$, T.\,Yamada$^2$, R.\,J.\,Ivison$^4$, T.\,Hayashino$^5$, K.\,Ohta$^6$, Y.\,Shioya$^7$ and Y.\,Taniguchi$^7$}
\vspace*{6pt}\\
\noindent $^1$Department of Physics, University of Durham, South Road, Durham. DH1 3LE. U.K.\\
\noindent $^2$National Astronomical Observatory of Japan, Mitaka, Tokyo, 181-8588. Japan\\
\noindent $^3$Department of Physics, California Institute of Technology, MS 320-47, Pasadena, CA. 91125. USA\\
\noindent $^4$Astronomy Technology Centre, Royal Observatory, Blackford Hill, Edinburgh. EH9 3H5. U.K.\\
\noindent $^5$Research Center for Neutrino Science, Graduate School of
Science, Tohoku University, Aramaki, Aoba, Sendai 980-8578, Japan \\
\noindent $^6$Department of Astronomy, Kyoto University, Sakyo-ku, Kyoto
606-8502. Japan\\
\noindent $^7$Astronomical Institute, Graduate School of Science, Tohoku
University, Aramaki, Aoba, Sendai 980-8578. Japan }
 

\date{}

\pagerange{\pageref{firstpage}--\pageref{lastpage}} \pubyear{2005}

\maketitle

\label{firstpage}

\begin{abstract}
We present the results from a submillimetre (submm) survey of a sample of 23
giant Ly$\alpha$ emitting nebulae in the overdensity at $z=3.09$
in the SA\,22 field. These objects, which have become known as Ly$\alpha$ Blobs
(LABs), have a diverse range of morphology and surface brightness, but
the nature of their power-source remains unclear, with both cooling flows or
starburst/AGN ionised winds being possibilities. Using the SCUBA submm
camera on the JCMT, we measure the 850\,$\mu$m
flux of a sample of LABs.  We present detections of
submm emission from four LABs at $>3.5\sigma$
individually, and obtain a modest statistical detection of the full sample at
an average flux of $3.0\pm0.9$\,mJy.  These detections indicate
significant activity within the LAB haloes, with bolometric
luminosities in the ultraluminous regime ($>10^{12}$\,$L_\odot$),
equivalent to a star formation rate of
$\sim$$10^3$\,$M_\odot$\,yr$^{-1}$. By comparisons to LAB-like objects in other regions, we show that there is an apparent trend (although weak)
between observed Ly$\alpha$ emission and bolometric
luminosity. Combined with 
our detection of ultraluminous activity in this population and
the lack of any strong  morphological correlations in our sample, 
this provides evidence that the interaction of an ambient halo of
gas with a galactic-scale ``superwind'' is most likely 
to be responsible for the extended
Ly$\alpha$ emission in the majority of LABs. 
Assuming the extent of the LABs reflects outflows from a superwind,
we estimate the age of starbursts in the submm LABs to be in the range
10--100\,Myr. Using the average submm flux of the LABs, we
determine a star-formation rate density in the SA\,22 structure 
of $>$3\,$M_\odot$\,yr$^{-1}$\,Mpc$^{-3}$, greater than the field at this 
epoch.  The submm detection
of these four LABs means there are now 7 luminous submm galaxies in the  $z=3.09$ structure in SA\,22, making this the
largest known association of these intensely active galaxies. 
This clustering further strengthens the
proposed evolutionary link between these galaxies and local
cluster ellipticals.  Finally we suggest
that the highly-extended Ly$\alpha$ haloes (which define the LAB class) 
may be a common feature of the submm galaxy
population in general, underlining their role as potentially important
sources of metal enrichment and heating of the intergalactic medium.
\end{abstract}


\section{Introduction}
\label{sec:intro}

%
%
\begin{table*}
\caption{The catalog of LABs  in the SA\,22 region observed in the
  submm.
We give the
coordinates, 850\,$\mu$m fluxes, isophotal Ly$\alpha$ emission areas and Ly$\alpha$
luminosities for LABs in the full sample. LABs detected at $>3.5\sigma$ 
significance at 850\,$\mu$m are shown in bold face type.
For comparison, we also note the
results for the well studied LAB1 and LAB2. We also classify the
objects based on a simple morphological/Ly$\alpha$ luminosity
description: F/C (faint+compact), F/E (faint+extended), B/C
(bright+compact) and B/E (bright+extended), see \S\ref{ssec:submm}. The compact/extended
boundary is 50 arcsec$^2$ (2900 kpc$^2$), and the faint/bright boundary is
$10^{43}$\,ergs\,s$^{-1}$. Note: 10$^{44}$ ergs\,s$^{-1}$ = $2.6\times10^{10}$ $L_{\odot}$.
}
\label{tab:labs}
\begin{tabular}[h]{lccr@{$\pm$}lccr@{.}lccl}
\hline
\multirow{2}{1cm}{Name} & RA (J2000) & Dec. (J2000) & \multicolumn{2}{c}{$S_{850}$}     & Area$^a$              & $\log_{10} L_{\rm Ly\alpha}^b$ & \multicolumn{2}{c}{$\log_{10} L_{\rm bol}^c$}  & Luminosity/ &  \multirow{2}{1cm}{Notes$^d$}\\
     & (h m s)   &  (d $'$ $''$) & \multicolumn{2}{c}{(mJy)}         & (arcsec$^2$)          & (ergs\,s$^{-1}$)               &  \multicolumn{2}{c}{(ergs\,s$^{-1}$)}      & morphology &&  \\
\hline

{\bf LAB1} & {\bf 22 17 24.68} & {\bf+00 12 42.0} &  {\bf 16.8}& {\bf 2.9}           & {\bf 222}   & {\bf 44.04 } & {\bf 47}&{\bf 06} & {\bf B/E} & {\bf 2 companion LBGs}\\
     LAB2  &      22 17 39.00  &     +00 13 27.5  &       3.3  &      1.2            &      152    &     43.93    &     $<$46&39  &    B/E  &      Hard X-ray source\\
\hline

LAB3       & 22 17 59.15        & +00 15 29.1 		&      $-$0.2&1.5  		& 78 		& 43.76 	& $<$46&48  & B/E 		& Ly$\alpha$ Emitter (LAE)$^{\dagger}$\\
LAB4 	   & 22 17 25.12        & +00 22 11.2		&      0.9&1.5  		& 57 		& 43.58 	& $<$46&49 		& B/E 		& \\
{\bf LAB5} & {\bf 22 17 11.67}  & {\bf+00 16 44.9} 	&     {\bf5.2}&{\bf1.4}  	& {\bf55}	& {\bf43.23} 	&{\bf 46}&{\bf 55} 	& {\bf B/E} 	& {\bf LAE}\\
LAB6 	   & 22 16 51.42 	& +00 25 03.6 		&      $-$0.5&1.8  		& 42 		& 43.20 	& $<$46&56 		& B/C 		& LAE\\
LAB7       & 22 17 40.99 	& +00 11 26.9 		&      0.2&1.6  		& 40 		& 43.18 	& $<$46&51 		& B/C\\
LAB8       & 22 17 26.18 	& +00 12 53.5  		&      0.3&5.3 			& 39 		& 43.23 	& \multicolumn{2}{}{} 		& B/C 		& {\footnotesize MAP}\\
LAB9       & 22 17 51.09 	& +00 17 26.2		&      1.3&5.3			& 38 		& 43.11 	& \multicolumn{2}{l}{} 		& B/C 		& LAE, {\footnotesize MAP}\\
{\bf LAB10}& {\bf 22 18 02.27}  & {\bf +00 25 56.9}	&      {\bf 6.1}&{\bf1.4}  	& {\bf34}	& {\bf43.34} 	&  {\bf 46}&{\bf 61}  &{\bf B/C} 	& {\bf LAE}\\
LAB11      & 22 17 20.33 	& +00 17 32.1		&      $-$0.4&5.3		& 30 		& 42.96 	& \multicolumn{2}{c}{}		& F/C  		& {\footnotesize MAP}\\
LAB12      & 22 17 31.90 	& +00 16 58.0 		&      3.2&1.6 		& 29 		& 42.93 	& $<$46&52		& F/C 		& \\
{\bf LAB14}&{\bf 22 17 35.91}	& {\bf +00 15 58.9}	&      {\bf 4.9}&{\bf 1.3}	& {\bf 27} 	& {\bf 43.08}  	&{\bf 46}&{\bf 52 }  & {\bf B/C} 	& {\bf LAE, C05} \\
LAB16      & 22 17 24.84 	& +00 11 16.7  		&      2.2&5.3			& 25		& 43.00 	& \multicolumn{2}{c}{}		& B/C		& LAE, {\footnotesize MAP}\\
{\bf LAB18}& {\bf 22 17 28.90 }	& {\bf +00 07 51.0}	&      {\bf 11.0}&{\bf 1.5}  	& {\bf 22 }	& {\bf 42.81}	& {\bf 46}&{\bf 87} 	& {\bf F/C}&Possible X-ray source\\
LAB19      & 22 17 19.58 	& +00 18 46.5 		&      $-$8.6&5.3		& 21 		& 43.11 	& \multicolumn{2}{c}{} 		&B/C 		& LAE, {\footnotesize MAP}\\
LAB20      & 22 17 35.30 	& +00 12 49.0 		&      0.4&1.5  		& 21 		& 42.81 	& $<$46&49 		& F/C\\
LAB25      & 22 17 22.59 	& +00 15 50.8  		&      1.4&5.3 			& 19 		& 42.77 	& \multicolumn{2}{c}{} 		& F/C 		&{\footnotesize MAP}\\
LAB26      & 22 17 50.43 	& +00 17 33.4  		&      $-$2.7&5.3 		& 18 		& 42.79 	& \multicolumn{2}{c}{} 		& F/C 		& {\footnotesize MAP}\\
LAB27      & 22 17 06.96 	& +00 21 31.1		&      0.5&1.6  		& 18 		& 42.83 	& $<$46&50		& F/C\\
LAB30      & 22 17 32.44 	& +00 11 34.1		&      3.3&1.3  		& 17 		& 42.98 	& $<$46&41	 	& F/C 		& \\
LAB31      & 22 17 38.94 	& +00 11 02.0  		&      $-$3.7&5.3 		& 17 		& 43.04 	& \multicolumn{2}{c}{}		& B/C 		& LAE, {\footnotesize MAP}\\
LAB32      & 22 17 23.88 	& +00 21 56.5		&      1.8&1.4  	 	& 17 		& 42.76 	& $<$46&46 	  	& F/C 		& LAE\\
LAB33      & 22 18 12.56 	& +00 14 33.3		&      1.6&1.5  		& 16 		& 42.95 	& $<$46&49 		& F/C 		& \\
LAB35      & 22 17 24.84 	& +00 17 17.0 		&      1.2&5.3			& 16 		& 42.95  	& \multicolumn{2}{c}{} 		&F/C 		&LAE, {\footnotesize MAP}\\
\hline
\multicolumn{9}{l}{$^a$ M04 -- isophotal area determined on corrected narrow-band emission-line map}\\ 
\multicolumn{9}{l}{$^b$ M04 -- luminosity at $z=3.09$} \\
\multicolumn{11}{l}{$^c$ bolometric luminosity determined from $S_{850}$, assuming a modified black-body with $T_{\rm d} = 40$\,K, $\alpha=4.5$, $\beta=1.7$.}\\
\multicolumn{12}{l}{$^d$ taken from the literature: S00, M04, Hayashino et al.\ (2004), Chapman et al.\ (2005). MAP $=$ SCUBA scan map extracted flux, see \S\ref{ssec:photom}} \\
\multicolumn{9}{l}{$^\dagger$ Associated optical Ly$\alpha$ emitter from Hayashino et al.\ (2004).}\\ 

\end{tabular}
\end{table*}
  
Since the earliest X-ray spectroscopic observations of clusters of
galaxies, it has been apparent that the intracluster medium (ICM)
in these systems contains a significant mass of metals.  This
material must have been processed through stars (most likely residing in
galaxies) and then either
expelled or removed from the stellar system. The apparent lack
of evolution in the ICM metallicity in the high density
cores of clusters out to $z\sim 1$ (Tozzi et al.\ 2003;
Mushotzky \& Scharf 1997) points to
an early phase of ICM enrichment for the
highest density regions, while claims of a minimum entropy
in the ICM (Ponman et al.\ 1999) 
support mechanisms which can transfer both energy and metals
to the environment. Outflows or winds driven by star formation
or AGN activity have been proposed 
as an  efficient mechanism for dispersing metals from
within galaxies and heating the surrounding gas. 
Local starburst galaxies such as M82 or Arp\,220
are known to exhibit spatially-extended structures, 
visible as haloes of emission
line gas and hot X-ray emitting gas.  These are believed to be
related to wind-driven
outflows emerging from the active regions of these
galaxies (Heckman et al.\ 1990). If high redshift
galaxies are expelling considerable quantities of enriched
material into their environments then 
similar emission-line structures may be visible around them.

Sensitive narrow-band surveys have uncovered a class of giant
Ly$\alpha$ emission-line nebulae at high redshift (Steidel et al.\ 2000 (S00); Matsuda et al.\ 2004 (M04)). These surveys both focus on an overdense
structure at $z=3.09$ discovered by S00 in the SA\,22 field.
This structure has been
interpreted as a cluster in the process of formation
(S00), and is therefore an excellent natural laboratory to study galaxy
evolution. The extended and morphologically
diverse features of the Ly$\alpha$
haloes in SA\,22 have led them to be termed Ly$\alpha$ Blobs (LABs, S00). 
With luminosities of up to $10^{44}$\,ergs\,s$^{-1}$ and physical
extents of up to 100\,kpc (S00), these objects have many of the
properties expected for high-redshift analogs of the outflowing haloes seen
in local galaxies -- although their luminosities and sizes would need to be scaled up
by at least an order of magnitude to match the high-redshift LABs. 

Similarly extended and luminous Ly$\alpha$ halos have been found around         
bright radio galaxies at comparable redshifts using similar narrow-band         
imaging techniques (e.g.\ De Breuk et al.\ 1999; Kurk et al.\ 2000;             
Reuland et al.\ 2003).  These halos are interpreted as tracers of               
cooling and feedback in merging galaxies at high redshifts (Reuland et          
al.\ 2003), with their morphological diversity suggesting the active            
growth through mergers of these powerful radio galaxies.  There is also         
strong evidence that many of these radio galaxies reside in                     
high-density regions, as traced by overdensities of extremely red               
objects, Lyman-$\alpha$ emitters, X-ray sources or submillimetre                
galaxies (e.g.\ Kurk et al.\ 2000; Smail et al.\ 2003; Stevens et al.\          
2003).  Hence there is some circumstantial evidence of a connection             
between large Lyman-$\alpha$ halos and the earliest formation phase of          
the most massive galaxies in high-density regions at high redshifts.       

M04 present a new catalogue of 35 LAB candidates over a 0.5-degree field
at $z=3.09$ in the SA\,22 field from a narrow-band imaging programme using
Suprime-Cam on the 8.2-m Subaru Telescope.
This survey shows that LABs come in a
range of sizes ($\sim$ 20--200\,arcsec$^2$, $\sim$1150--11500 kpc$^2$) and have Ly$\alpha$
luminosities of the order $10^{42-44}$\,ergs\,s$^{-1}$. Their
morphologies are complex, including filamentary structures, apparent bubbles
and shells. Some LABs also appear to have associated continuum sources
which may be capable of providing sufficient ultra-violet photons to power
the emission line halo, but some do not.  Clearly several possible mechanisms could produce a LAB, not just
a wind, and several explanations of the physical
processes governing the Ly$\alpha$ emission have been proposed: (a)
photo-ionisation by massive stars or an obscured AGN; (b) cooling
radiation from a collapsing gaseous halo (Fardal et al.\ 2001) or (c)
starburst superwind shock heating (Taniguchi \& Shioya 2000, Ohyama et al.\ 2003). Additionally, it has been suggested that inverse-Compton scattering of cosmic microwave background photons by a population of relativistic electrons could also contribute (Scharf et al.\ 2004). 

While each of
these scenarios are potentially viable, one third of LABs are not
associated with ultra-violet (UV) continuum sources luminous enough to
produce Ly$\alpha$ emission from photo-ionization (assuming a Salpeter
initial mass function, M04), suggesting that the photo-ionization source
in (a) would additionally need to be  heavily obscured for
these LABs, indicative of dust in the system. This is
intriguing, since dust heavily suppresses Ly$\alpha$ emission,
requiring that the Ly$\alpha$ emission originates well away from the
obscuration, or that there is some other mechanism allowing Ly$\alpha$
photons to escape through the dust and into the observer's line of
sight.  The nature of the processes responsible for producing the
extended Ly$\alpha$ emission in LABs thus remains uncertain.

The first LAB to be studied in detail was LAB1 in the SA\,22
field (S00; Chapman et al.\ 2001 (C01); Chapman et al.\ 2004 (C04)).  
LAB1 is the most luminous LAB cataloged in the SA\,22 structure 
with a Ly$\alpha$ luminosity of $L_{\rm Ly\alpha}
 = 1.1\times10^{44}$\,ergs\,s$^{-1}$ and also has the largest extent of all LABs 
identified to date, at $\sim$100\,kpc. LAB1's morphology is complex, and 
its includes companion galaxies and other structures
visible at high resolution 
in {\it Hubble Space Telescope (HST)} imagery from C04. 
There is now a wealth of multi-wavelength data
available for this object.  In particular,
C01 show that there is a strong submillimetre source
coincident with the nebula ($S_{850} = 16.8\pm2.9$\,mJy),
confirming the presence of an extremely luminous power source within 
LAB1 with a bolometric luminosity in excess of 10$^{13}L_\odot$. 
C04 note that deep {\it Chandra} X-ray
observations  in the region of LAB1 failed to detect
an X-ray counterpart -- suggesting the bolometric emission
is not powered by an unobscured or partially obscured luminous
AGN. However, they suggest that a heavily obscured
AGN with a torus orientated at 45$^\circ$ to the sky might be
responsible for the Ly$\alpha$ halo and would also 
explain extended linear features revealed by {\it HST} imaging, suggestive of
jet induced star-formation.

Bower et al.\ (2004) used the SAURON integral field
unit (IFU) to map the 2-dimensional dynamics
of the Ly$\alpha$ emission in LAB1, including the haloes
of two associated Lyman Break
Galaxies (LBGs, C11 and C15). These authors conclude that the nebula
has a complex velocity structure which cannot be explained by a simple
shell-like outflow; and that the submm source occupies a cavity in the
Ly$\alpha$ halo, suggesting that either the region in the immediate
vicinity of the SMG is obscured by dust ejecta, or it has completely ionised the material in this region.  

The local surface density of 283 Ly$\alpha$ emitters (smoothed with a Gaussian with a kernal of $\sigma = 1.5$-arcmin) at LAB1 is 0.63\,arcmin$^{-2}$ compared to the average density over the whole Suprime-Cam field-of-view of 0.39\,arcmin$^{-2}$ (Hayashino et al.\,2004, M04). At a shallower limit, the density of the SA\,22  Suprime-Cam field is 1.75$\times$ that of the other blank (SXDS) field in Hayashino et al. (2004). Thus, the local density at LAB1 is 2.8$\times$ that of the blank field. Since LAB1 is located at a peak in the underlying surface density, perhaps it is not surprising that this object is
the brightest and largest of all known LABs -- as it could 
represent a massive
galaxy in the process of formation. 
It is equally possible that the large extent
and complicated velocity structure of LAB1
is actually generated from several overlapping haloes
related to C11, C15 and the submm source, which happen to
inhabit the same dense region in the $z=3.09$ structure.

The other LAB identified by S00 in the SA\,22 structure is LAB2, 
this has a comparable Ly$\alpha$ luminosity to
LAB1 and  it has recently been shown to contain a hard X-ray
source (Basu-Zych \& Scharf 2004). These authors suggest that if the X-ray source
is point-like, then the unabsorbed X-ray luminosity of 
$L_X\sim10^{44}$\,ergs\,s$^{-1}$ indicates that LAB2 harbours
a super-massive black hole with a high local absorbing column.
C01 also report submm emission from the vicinity of
LAB2,  but with a significantly lower submm flux than LAB1: $S_{850} =
3.3\pm1.2$\,mJy (C01; C04). The detection of  submm emission from both of the LABs studied to date
may suggest a strong link between these two populations.  This would
clearly favour those models for the formation of LABs which rely upon
a highly active source to generate the extended emission-line halo.
Perhaps more
interestingly, if some LABs do not contain bolometrically-luminous sources, then we must
consider a variety of processes (cooling, photo-ionisation, etc.) to
account for these giant haloes. We have therefore undertaken a survey
to detect or place limits on the submm emission from LABs at
$z=3.09$ in the SA\,22 field.

\medskip

\noindent The paper is organised as follows: \S\ref{sec:obs}
describes the LAB sample and data reduction, \S\ref{sec:anal} contains our analysis, in \S\ref{sec:dis} we discuss the results and in
\S\ref{sec:con} we report  our conclusions. Unless otherwise stated, we
adopt a flat Universe with $\Omega_m=0.3$, $\Omega_\Lambda=0.7$ and
$H_0 = 70h$ (km\,s$^{-1}$\,Mpc$^{-1}$), with $h=1$. In this cosmology
the angular scale at $z=3.09$ is 7.6\,kpc arcsec$^{-1}$ and the look-back
time is 11.4\,Gyr.

\begin{figure}
\begin{center}
\begin{turn}{-90}
\includegraphics[scale=0.47]{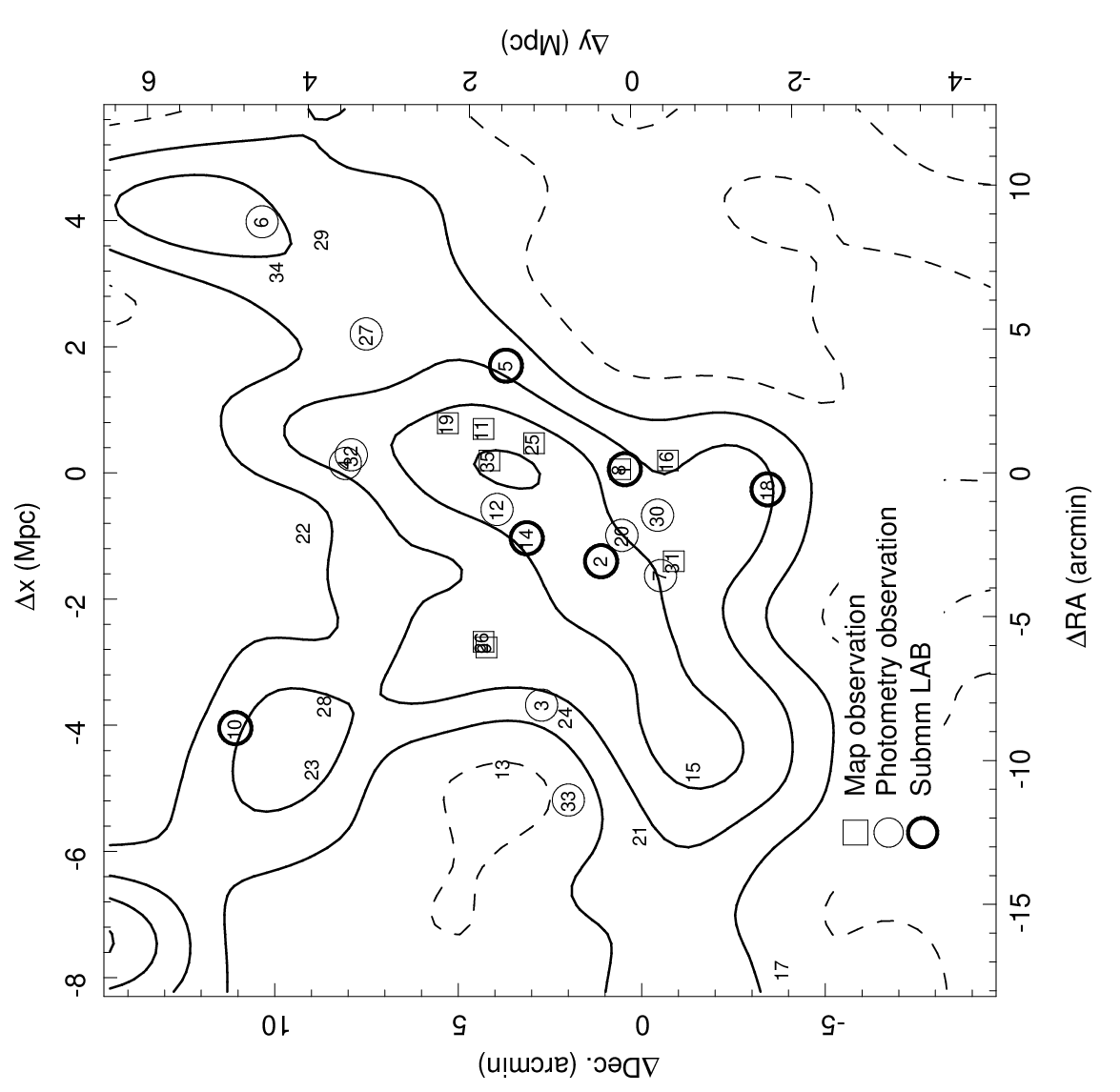}\end{turn}
\end{center}
\caption{The sky-distribution of the 35 known LABs in SA\,22 from M04. Photometry observations are circled, map observations are identified with squares and submm detected LABs are circled in bold (\S\ref{ssec:submm}). The contours represent the surface density of the underlying LAE distribution: solid(dashed) contours show regions of greater(less) than average surface density in steps of 50\% of the average surface density. We discuss the enviroments of LABs further in \S\ref{ssec:env}. The origin of the coordinates are centred on LAB1, and the physical scale corresponds to angular separation.}
\label{fig:skyplot}
\end{figure}

\section{Observations \& Data Reduction}
\label{sec:obs}

Our selection was based on Subaru Suprime-Cam narrow-band imaging of a
$34 \times 27$ arcmin region around SA\,22, which detected 33 new
LABs (M04), in addition to the two LABs previously catalogued by S00.
M04 obtained deep narrow- and broad-band imaging of this region using a
narrow-band filter,
NB497, and {\it BVR} filters with Suprime-Cam. 
The central wavelength of NB497 is 4977\AA,
with a FWHM of 77\AA, sensitive to Ly$\alpha$ emission over a
redshift range of $z=3.06$--3.13. By combining the $B$ and $V$
bands they were able to estimate the 
continuum emission in the narrow-band filter. 
This allows construction of a continuum-subtracted, emission-line image
by subtracting the \emph{BV} image from the NB497 image. 
In this paper we adopt the
nomenclature of M04, who catalog the 35 LABs in order of descending
isophotal area. In this system LAB1 and LAB2 correspond to the ``blob 1''
and ``blob 2'' of earlier works (e.g.\ S00; C01).

We selected 13 of these LABs as targets for our submm observations.
These were chosen to cover the full range
of properties spanned by the LAB population (area, brightness, 
morphology, environment, see Table~1 and Fig.~\ref{fig:skyplot}).
To these 13 we have identified a further  SMG,
SMM\,J221735.84+001558.9, a submm source
catalogued by Barger, Cowie \& Sanders (1999) which corresponds to a
Ly$\alpha$-emitting $\mu$Jy radio source at $z=3.09$ 
from Chapman et al.\ (2005).  This SMG is coincident with the
extended Ly$\alpha$ halo, LAB14, from M04.  We note that Chapman
et al.\ (2005) have detected a further SMG at $z=3.098$ within the
overdensity, SMM\,J221735.15+001537.2, with $S_{850}=6.3\pm1.3$\,mJy
although this SMG has detectable Ly$\alpha$
emission this is not 
sufficiently spatially extended to class as an LAB based on
the selection of M04.

We used the National Radio Astronomy Observatory's\footnote{NRAO is
operated by Associated Universities Inc., under a cooperative agreement
with the National Science Foundation.} (NRAO)  Very Large Array (VLA) to target SA\,22 during October 1998 and July--September 2003, obtaining 48\,hr of data at 1.4\,GHz. The observations were taken for approximately 12\,hr in B configuration and the remainder in A
configuration. The pseudo-continuum correlator mode (`4') was employed,
with 28 $\times$ 3.25-MHz channels, enabling us to map almost the entire
primary beam ($\rm 27.3'\,\times\,27.3'$), recording data every 5\,sec in
both left-circular and right-circular polarisations. 0137+331 was used to
set the flux scale, with 2212+018 (2.8\,Jy) used for local
phase/amplitude/bandpass calibration.

Finally, in addition to our new 
submm photometry observations, we also take advantage of an existing shallow submm map to place limits on a further 9 LABs. In total, including the previous published studies of LAB1 \& 2, we have submm observations of 25 LABs.

\subsection{Submm observations}
\label{ssec:photom}

Observations were conducted using the Submillimetre Common User
Bolometer Array (SCUBA) on the James Clerk Maxwell Telescope (JCMT)
over the nights of 2004 September 18--22. SCUBA was used in photometry
mode to search for 850\,$\mu$m and 450\,$\mu$m emission from the 13 LABs listed in Table~1.   
We employed two-bolometer chopping, whereby the on- and off-source
positions are 
divided between three bolometers  to maximise the signal-to-noise
ratio (SNR) in the final coadded flux measurement (discussed further in
\S\ref{sec:dr}).  Our goal was a mean r.m.s. noise in our measurements
of $\sim$1.5\,mJy -- requiring $\sim$2--3\,ks integration in Grade 1--2
weather conditions ($\tau_{\rm CSO}\leq 0.08$, $\tau_{850}\leq 0.32$).

To achieve sky and background cancellation we chopped in
azimuth by 60$''$.  Calibration observations employed 
Uranus and zenith opacity was measured from regular skydips and the
JCMT water vapour monitor, yielding $\tau_{850}\leq 0.25$ for
all observations, with $\tau_{850}\sim 0.1$ for some portions of
the run.  The on-sky exposure times were 2.2\,ks for all sources, which in
the conditions we experienced yielded mean 1-$\sigma$ noise limits of
1.5\,mJy and 13\,mJy at 850 and 450\,$\mu$m respectively.

We also obtained $S_{850}$ fluxes or limits for 
ten additional LABs (LAB8, 9, 11, 
16, 19, 25, 26, 31, 35, see Table~\ref{tab:labs}) which were not observed in photometry-mode,
but do fall within a shallow, submm scan-map of SA\,22 (Chapman \& Borys,
priv.\ comm.). The map data were taken in
a combination of SCUBA jiggle map and SCUBA raster map
modes at 850~$\mu$m during a number of observing runs
with good observing conditions ($\tau_{850}<0.09$). Some of the SCUBA jiggle maps are described in detail in
Barger, Cowie, \& Sanders (1999) and Chapman et al.\
(2001, 2003, 2004).
The combination of these SCUBA jiggle and raster maps are described in
Chapman et al.\ (2003), using the approach of Borys et al (2003). This map covers an $11\times 11$ arcmin 
region approximately
centred on LAB1 and has a conservative r.m.s.\ depth of $\sim$5.3\,mJy at
850\,$\mu$m -- it
therefore is only sensitive enough to detect the brightest submm sources,
such as that residing in LAB1. We extract the fluxes from the 
calibrated map by measuring 15$''$-diameter aperture fluxes and converting these
using an assumed Gaussian beam profile to give fluxes in Jy\,beam$^{-1}$. In Figure~\ref{fig:skyplot} we plot the sky distribution of all 35 LABs from M04, indentifying those which are map or photometry observations.

\subsection{Data reduction}
\label{sec:dr}

Standard routines from the SCUBA User Reduction Facility
({\sc surf}, Jenness \& Lightfoot 1998) were used to reduce each LAB
photometry observation. For both long and short arrays (850\,$\mu$m and
450\,$\mu$m), demodulated data were flat-fielded and corrected for
atmospheric extinction. A
sky-estimate was subtracted using {\sc remsky} by subtracting
the average (median) signal from all the off-source bolometers in
the arrays (avoiding any dead or noisy bolometers). Subtracting the sky gave
a marginally better SNR than not removing it. Finally, a 6-$\sigma$ clip
was applied to remove spikes before the signals were coadded. For
calibration, we derived the flux-conversion-factor (FCF) from the
observations of Uranus using the {\sc fluxes} program.

The chopping technique we adopted splits the
on-source integration over two bolometers in an attempt to maximise the exposure
time spent on the source. After the FCF was applied, an
inverse-variance weighting scheme was applied to combine the two
signals. Compared to the SNR from the main signal bolometer, on average
the SNR improved after combining by $\sim$14 per cent for $S_{850}$ and
$\sim$8 per cent for $S_{450}$. We list the fluxes measured for the LABs in Table~1. Note that the flux densities in Table~1 do not include uncertainties due to absolute flux calibration, which we estimate to be $\sim$10\%.

For the VLA data, editing and calibration was
performed using standard {\sc aips} procedures. We then employed {\sc
imagr} to map the central region, together with 40 satellite fields,
running several iterations of the self-calibration procedure described by
Ivison et al.\ (2002). Finally, we corrected for the primary beam response
using {\sc pbcor}.  The final image has a series of north-south stripes, as
described by C04, but this affects none of the LABs discussed here. The
typical noise level is around 9\,$\mu$Jy\,beam$^{-1}$, where the beam has a
{\sc fwhm} of 1.4$''$.

At the $\gtrsim\rm 4\sigma$ level, the only detections amongst our sample are
of LAB1 (as described by C04) and LAB4, which has several weak radio
sources near the position listed in Table~1, as well as a stronger source
to the south-east. The remainder of the LAB sample discussed
here have upper limits at 1.4\,GHz of around 5\,$\sigma<45\,\mu$Jy (for unresolved sources).

\section{Analysis \& Results}
\label{sec:anal}

\subsection{Submm emission}
\label{ssec:submm}

In Table~\ref{tab:labs} we summarise the results of our submm
observations, along with other information collected from the
literature. In addition to the photometry mode observations, we
tabulate the fluxes extracted from the SCUBA scan-map, and the
previously published
850\,$\mu$m fluxes of LAB1,\,2 from Chapman et al.\ (2000).   
We detect three LABs from the photometry-mode targets with
$S_{850}$ fluxes detected at significances
of $>3.5\sigma$: LAB5, 10 \& 18.   A further target, LAB30, has a flux
corresponding to $2.6\sigma$, but we do not classify this as a
formal detection.  No
individual LABs are detected at 450\,$\mu$m and therefore we have not
tabulated individual 450\,$\mu$m results.  

To these three new detections, we can also add a fourth submm-detected
LAB, LAB14.  This source was detected in jiggle-map observations
by Barger et al.\ (1999) with
an 850-$\mu$m flux of 4.9$\pm$1.3\,mJy (Chapman et al.\ 2005).  
The submm galaxy (SMG) was
included in the redshift survey of 
submm galaxies by Chapman et al.\ (2003, 2005).  They identified the
counterpart as a Ly$\alpha$-emitting galaxy at $z=3.089$, and
cross-correlation
of their position with the LAB catalog of M04 shows that this SMG
has an extended Ly$\alpha$ halo.  We can thus add four 
submm-detected LABs to the two previously known.

\begin{figure}

\begin{center}
\begin{turn}{-90}
\includegraphics[scale=0.47]{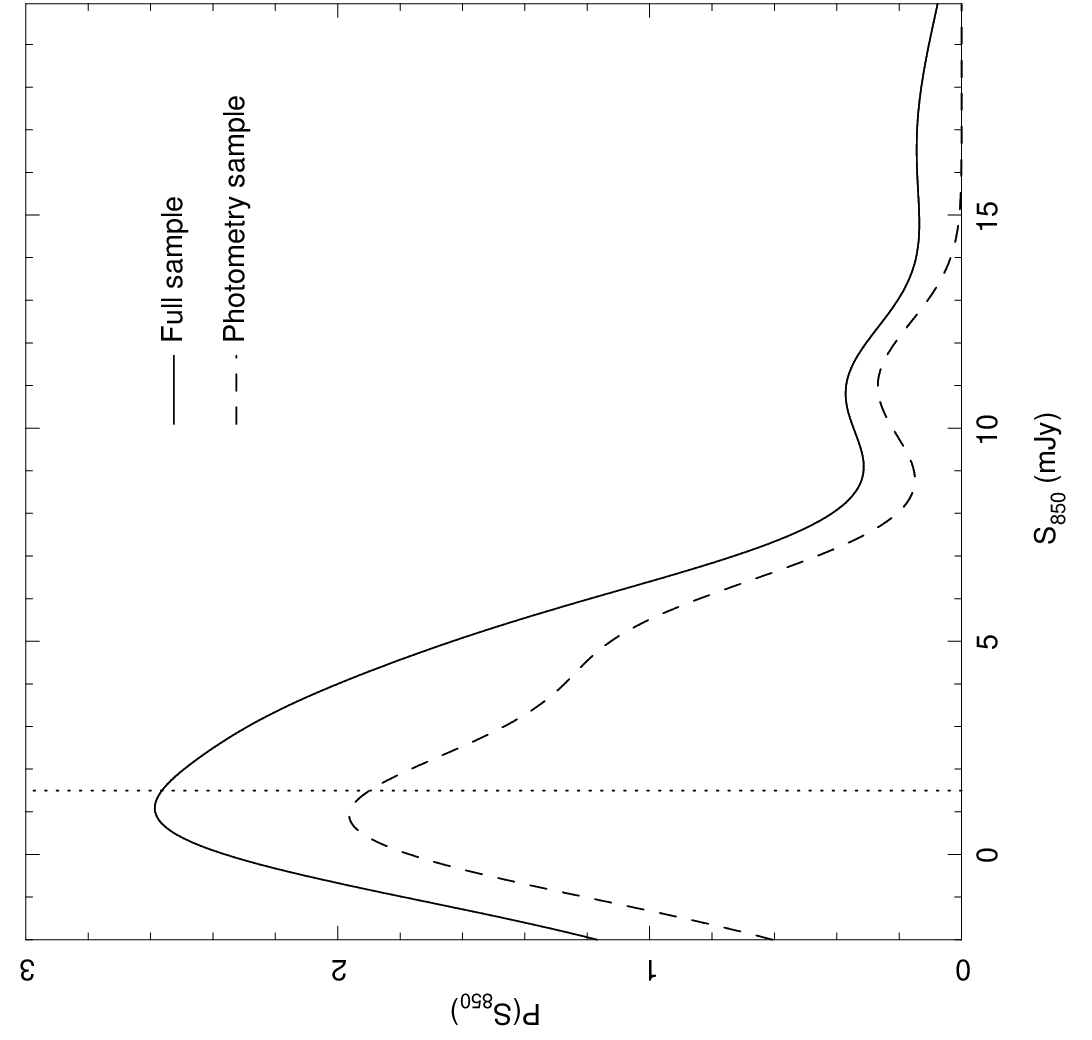}
\end{turn}
\end{center}
\caption{The 
$S_{850}$ flux distribution of LABs in the sample. Here we have represented the distribution in terms of a superposition of Gaussian probability distributions for the full sample (solid line) and just the photometry sample (dashed line). The photometry sample only includes those objects targeted with SCUBA in `photometry' mode, whereas the full sample includes LAB1, 2 and the map extractions. The vertical dotted line indicates the 1-$\sigma$ 850$\mu$m noise limit for the photometry sample.}
\label{fig:labhist}
\end{figure}

Looking at the distribution of 850 $\mu$m fluxes for the entire sample, 
Figure~\ref{fig:labhist}, we note that there is an excess of positive
flux measurements for our sample.  We confirm this by deriving a
noise-weighted average 850\,\,$\mu$m
flux for the LABs observed in photometry mode of $2.8\pm 0.9$\,mJy, where the
uncertainty is derived by bootstrap resampling.  A more rigorous
test is provided by removing the well-detected LABs, averaging
the fluxes of the remainder gives a mean flux of $1.2\pm 0.4$\,mJy.
Similarly, when LAB1, 2, 14 and the fluxes
of those sources covered by the scan-map  are included in
this analysis the average flux is $3.0\pm 0.9$\,mJy (we note that our bootstrap errors are conservative compared to the noise-weighted r.m.s. values).  Thus we
have $\geq3\sigma$ detections of the  LAB samples with  
typical fluxes around the level of the blank-field SCUBA confusion limit.

Performing a similar analysis on the 
450\,$\mu$m fluxes for the photometry sample yields
a noise-weighted average of $6.2\pm 2.1$\,mJy, giving a 3-$\sigma$
detection of the whole sample at 450\,$\mu$m.  The noise-weighted
mean 450/850$\mu$m flux ratio is $S_{450}/S_{850} = 3.1\pm 2.8$.
This is consistent with a dust temperature of $T_{\rm d}\sim40$\,K at
$z=3.09$ (assuming a dust emissivity index of $1.5$).

We can convert the average 850-$\mu$m flux of the LABs into a 
typical luminosity 
using the characteristic dust temperature of $T_{\rm d}\sim40$\,K
derived above and knowing that the LABs are at $z=3.09$.
We estimate the typical LAB (with a Ly$\alpha$
luminosity of $6\times 10^{42}$\,ergs\,s$^{-1}$,
or $1.6\times 10^{9}$\,$L_\odot$) has a bolometric luminosity of
$5.4\times 10^{12}$\,$L_\odot$.  If this emission arises wholly
from star-formation with a standard IMF (Kennicutt 1998),
then this luminosity corresponds to a star formation rate (SFR)
of $\sim$900\,$M_\odot$\,yr$^{-1}$.

We note that the lack of radio detections for the majority of LABs is consistent with these estimates. Emipirically scaling the SEDs of either Arp\ 220 or M82 to our average 850$\mu$m flux for a source at $z=3.1$ implies 1.4GHz fluxes of 15--35$\mu$Jy -- below our detection limit. 

\subsection{Individually detected LABs}
\label{ssec:ind}

Here we discuss the properties of the individual LABs from our
sample.
We show the Ly$\alpha$
emission-line images (from M04) of a
selection of the  LABs in our
photometry sample in Figure~\ref{fig:labplots1}. Note that the SCUBA beam size is 15$^{''}$.

\subsubsection{LAB5}

LAB5 is a large halo, the second largest submm-detected LAB after LAB1, but
much fainter and more diffuse than LAB1. 
There is some evidence for a slight elongation of
the Ly$\alpha$ emission (Figure~\ref{fig:labplots1}),
but the halo overall is roughly circular and the brightest region
corresponds to a faint continuum source in the centre of the halo. LAB5 is located near the peak in the surface density of LAEs (Figure~\ref{fig:skyplot}).

\subsubsection{LAB10}

LAB10 has a relatively compact morphology, and is within about 5 arcsec
of an extended, bright continuum source. It is unclear whether these
objects are related, but if they are independent, then the 
submm detection of LAB10 combined with the lack of an
extended diffuse halo offers several possibilities for its
power source: (a) LAB10 is in the initial stages of luminous activity
(i.e.\ the start of a starburst or AGN), and is expelling a wind which
has so-far only progressed a small distance into the IGM; (b) the fact
that LAB10 is compact is an orientation effect: we are looking face on to
a jet or collimated wind emerging from the SMG. A cooling flow cannot be
ruled out, but the fact that this LAB has a significant 850\,$\mu$m
flux (corresponding to a SFR of $\sim 1600$\,$M_\odot$\,yr$^{-1}$) is 
strong circumstantial evidence that the Ly$\alpha$ halo is related to the obscured activity in this source.

Whereas the majority of LABs reside close to the main peak in the underlying distribution of LAEs (see Fig.~\ref{fig:skyplot}), LAB10 lies in a secondary peak which resembles a filamentary structure to the north-west, connected to the main bulk of the proto-cluster.

\subsubsection{LAB14}

The submm source in LAB14 was identified from cross-correlating the LAB
 catalog in M04 with the redshift survey of SMGs (Chapman et al.\
 2005). Its morphology is compact, and comprises a bright component
 with a diffuse  $\sim 50$\,kpc-long extension to the south-east. The
source has a bolometric luminosity of $8.7\times 10^{12} L_{\odot}$, making this a ultraluminous class active galaxy. As with LAB5, LAB14 is located close to the peak of the surface density of LAEs.

\subsubsection{LAB18}

This is the strongest submm source in the photometry sample
($S_{850} = 11.0\pm1.5$\,mJy) and has an 
apparently hour-glass Ly$\alpha$ morphology. It is the second
brightest LAB, after LAB1, in terms of its submm emission.
It does not posess a
continuum counterpart, which suggests that either its ionising source is
completely dust-enshrouded, or that the power for the Ly$\alpha$ emission is
from a superwind.  Unlike LAB1, this object does not have companion LBGs,
which may contribute to the large measured extent of LAB1. LAB18's  Ly$\alpha$ luminosity is lower than LAB1. We note from Fig.~\ref{fig:skyplot} that LAB 18 appears to reside in a much lower density environment (reflected in the projected surface density of LAEs) compared to the other LABs.  

Using archival {\it XMM-Newton} data (PI: O.\ Almaini) of the SA\,22
field we tentatively detect a hard X-ray source at the location of
LAB18 with a (0.4--10 keV) flux of $\sim6
\times10^{-15}$\,ergs\,cm$^{-2}$ s$^{-1}$ at the $\sim2 \sigma$
level. The measured
flux is consistent with the X-ray flux of LAB2 (Basu-Zych \& Scharf
2003) and corresponds to $L_X\sim 5\times10^{44}$ ergs s$^{-1}$
if we assume a photon index of 1.5 and local absorbing column
$n_{\rm H} = 4.8\times10^{20}$ cm$^{-2}$ as did those authors. We do not
detect X-ray emission from any other LABs in this sample.  
If this detection is confirmed then it would 
suggest that LAB18 harbours a luminous, obscured AGN which
may contribute to the submm emission and 
which could also couple with the ambient gas
in the galaxy's halo.

\subsubsection{Non-detected SMGs}

Although our study hints at submm emission from LAB12 and 30
at the $\sim 2$--2.5-$\sigma$ level, the majority of the LABs we
targeted with our photometry observations were not detected. 
These  non-detections could be down to the uncertainty in where
the submm source lies within the LAB -- this is a particular concern
for the largest and most irregular LABs -- although the large size of
the SCUBA beam and the flux aperture we adopted means we are relatively insensitive to offsets of
less than $\sim 5$\,arcsec; or insufficient
sensitivity in our submm observations. The fact that we
obtain a 3-$\sigma$ detection of the stacked submm emission from the stacked
sources at a flux level of $\sim 3$\,mJy suggests that 
insufficient depth is likely to be a contributing cause
to the non-detection of all but the brightest LABs.  

%
%

%
%
\begin{table}
\caption{The mean properties of LABs detected and undetected at $S_{850} >
3.5\sigma$. We give the values for the LABs from our
photometry observations, and in ``[]'' include data from LAB1,2,14 and
the extracted fluxes from the map. The $S_{850}$ values are
noise-weighted averages and all errors are bootstrapped.}

\begin{center}
\begin{tabular}[h]{lccc}
\hline
 & ${\rm log}_{10} \left<L_{\rm Ly\alpha}\right>$ & $\left<\rm Area\right>$ & $\left<S_{850}\right>$ \\ 
 & (ergs\,s$^{-1}$) & (arcsec$^{2}$) & (mJy) \\
\hline
Detected    & 43.13$\pm$0.14& 37$\pm$8 & 7.3$\pm$1.5\\
  	    & [43.30$\pm$0.16] & [72$\pm$31] & [7.1$\pm$1.4] \\
Un-detected & 43.10$\pm$0.09 & 34$\pm$7  & 1.2$\pm$0.4\\
  	    & [43.09$\pm$0.08] & [36$\pm$7] & [1.4$\pm$0.4] \\
\end{tabular}
\end{center}
\label{tab:types}
\end{table}

%
%
\begin{table}
\begin{center}
\caption{850\,$\mu$m fluxes (in mJy) of LABs based on a simple
morphological-luminosity classification: the bright/faint boundary is
10$^{43}$\,ergs\,s$^{-1}$ and the extended/compact boundary is
50\,arcsec$^2$ (2900 kpc$^2$). 
We give values for the LABs from our photometry
observations, and in ``[]'' values
which include LAB1, 2, 14 and our extracted
map fluxes. The mean fluxes listed here are noise-weighted
and the errors are estimated from bootstrap resampling.}

\begin{tabular} {ccc}
\hline
         & Bright & Faint \\
\hline
Extended & 2.2$\pm$1.4 & ... \\
         & [3.3$\pm$1.4] & [...] \\
Compact  & 2.4$\pm$1.8 & 3.2$\pm$1.3 \\
         & [2.8$\pm$1.5] & [3.0$\pm$1.2] 
\end{tabular}
\end{center}
\label{tab:mb}
\end{table}

\subsection{Ly$\alpha$ properties of LABs}

We now search for correlations in the submm detection rate
and properties of LABs as a function of their Ly$\alpha$
properties to identify trends which might indicate the
nature and origin of their extended haloes.
Therefore, in Table~2 we compare the properties of detected ($>3.5\sigma$) and
undetected LABs. We do this both for those LABs in our photometry
sample and also for the full sample including LAB1, 2, 14 and the limits
from the submm map. 

The first point to note from Table~2 is that there appears to be
no difference in the  isophotal area of
the Ly$\alpha$ emission (measured from the
continuum-corrected narrow-band images) or Ly$\alpha$ luminosities between LABs with
detected submm emission and those without. Including LAB1 does affect
the area significantly, but since this object is the largest known, 
it could be considered a ``special''
object and not typical of LABs in general. 

We can also perform the reverse test, to determine the submm emission
from different samples of LABs differentiated by the extent and
brightness of their Ly$\alpha$ haloes.
We therefore provide a simple classification based
on their isophotal areas  and Ly$\alpha$ luminosities. First, we assign
four categories: bright-compact (B/C), bright-extended (B/E),
faint-compact (F/C) and faint-extended (F/E). The bright/faint boundary is
10$^{43}$\,ergs\,s$^{-1}$ and the extended/compact boundary is
50\,arcsec$^2$ (2800 kpc$^2$). 
The noise weighted average 850\,$\mu$m flux for each category is
presented in Table~3. Again, we perform this analysis for just the
photometry observations, and the full sample including LAB1,2,14 and map
observations. There are no F/E LABs in this scheme, 
in part due to the
fact that F/E LABs 
fall below the sensitivity limit of M04's narrow-band survey.
There appears to be no difference between any of these area/luminosity classes. 

\begin{figure*}

\begin{turn}{-90}\includegraphics[scale=0.2]{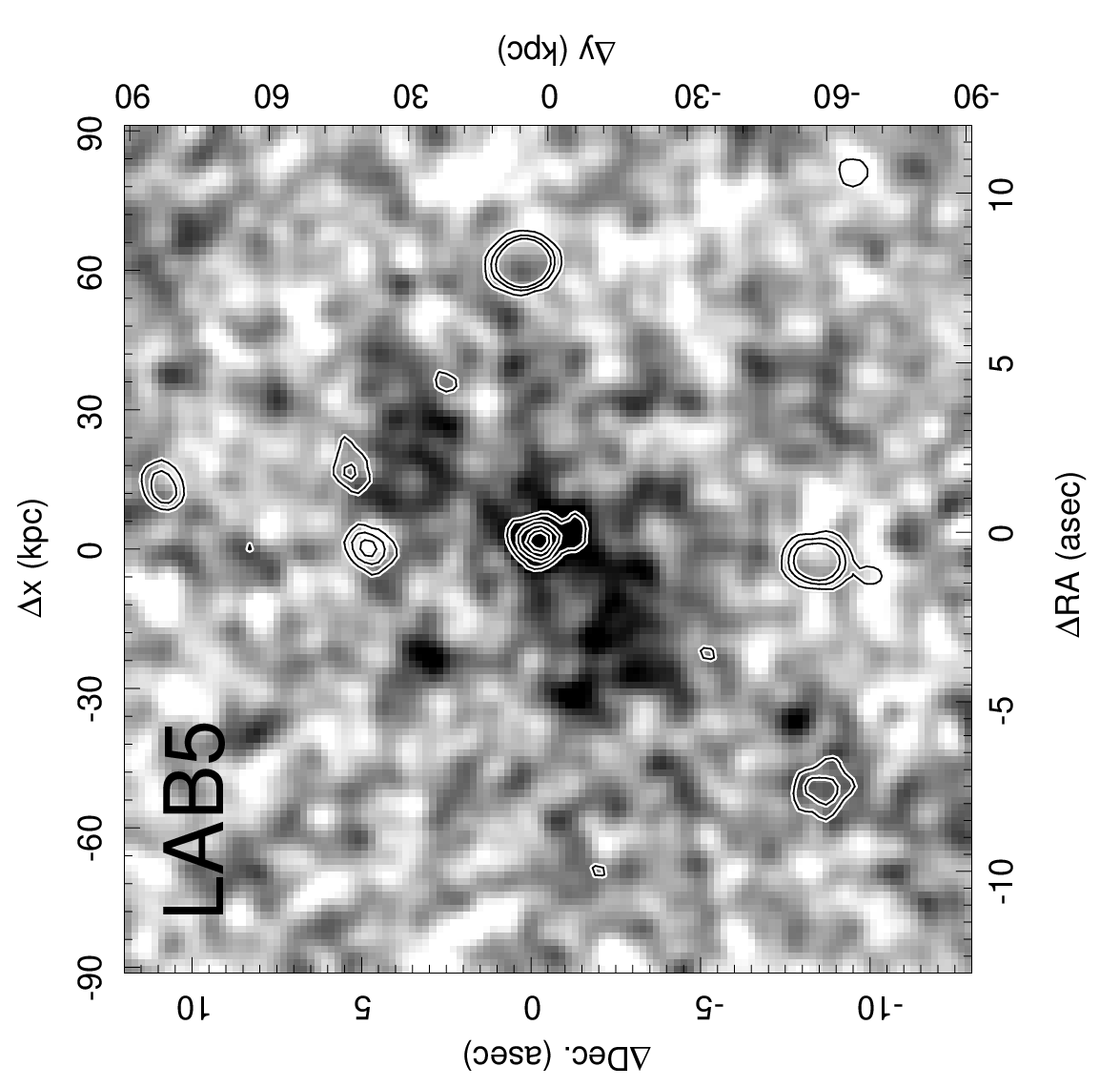}\end{turn}
\begin{turn}{-90}\includegraphics[scale=0.2]{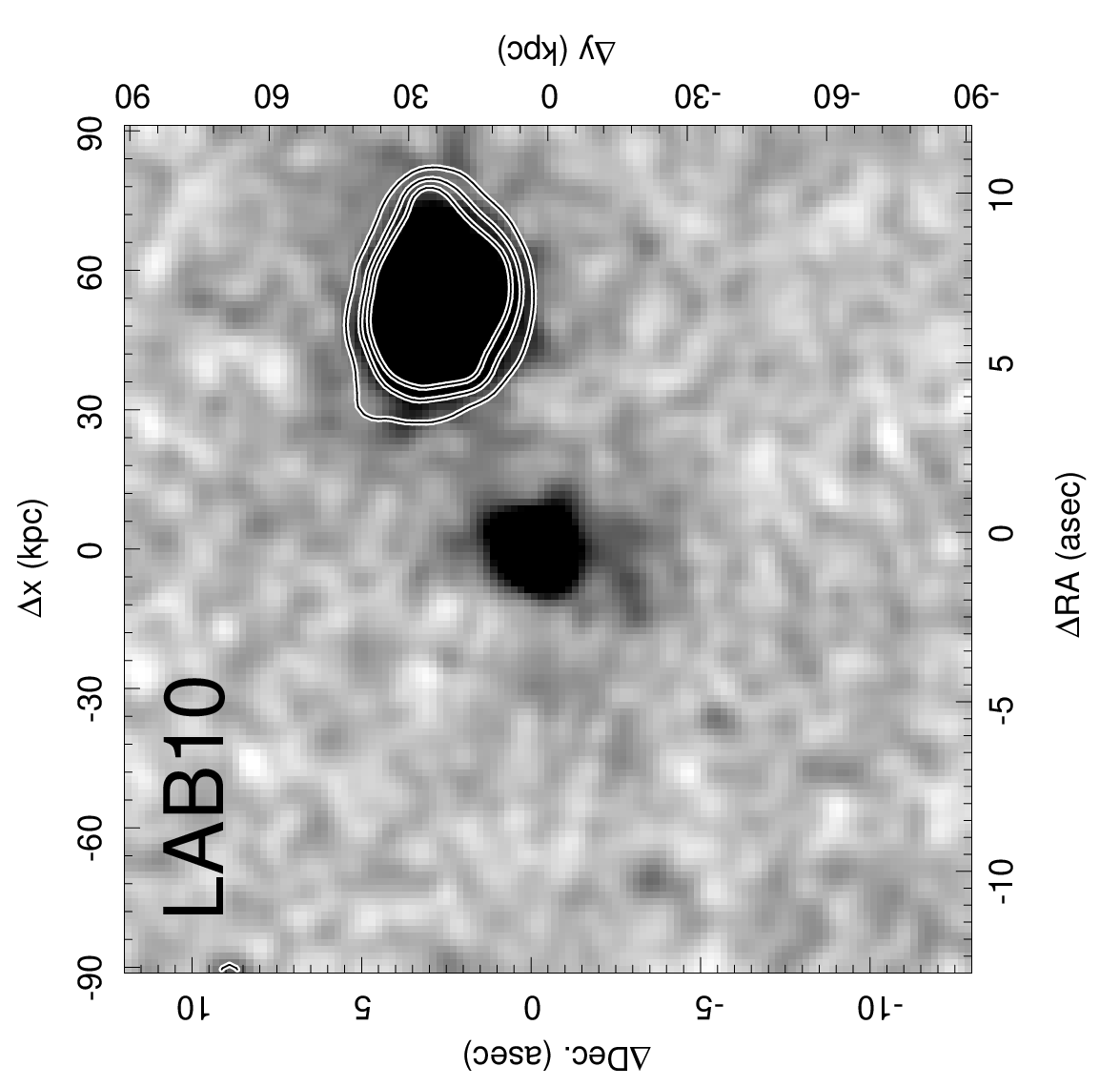}\end{turn}
\begin{turn}{-90}\includegraphics[scale=0.2]{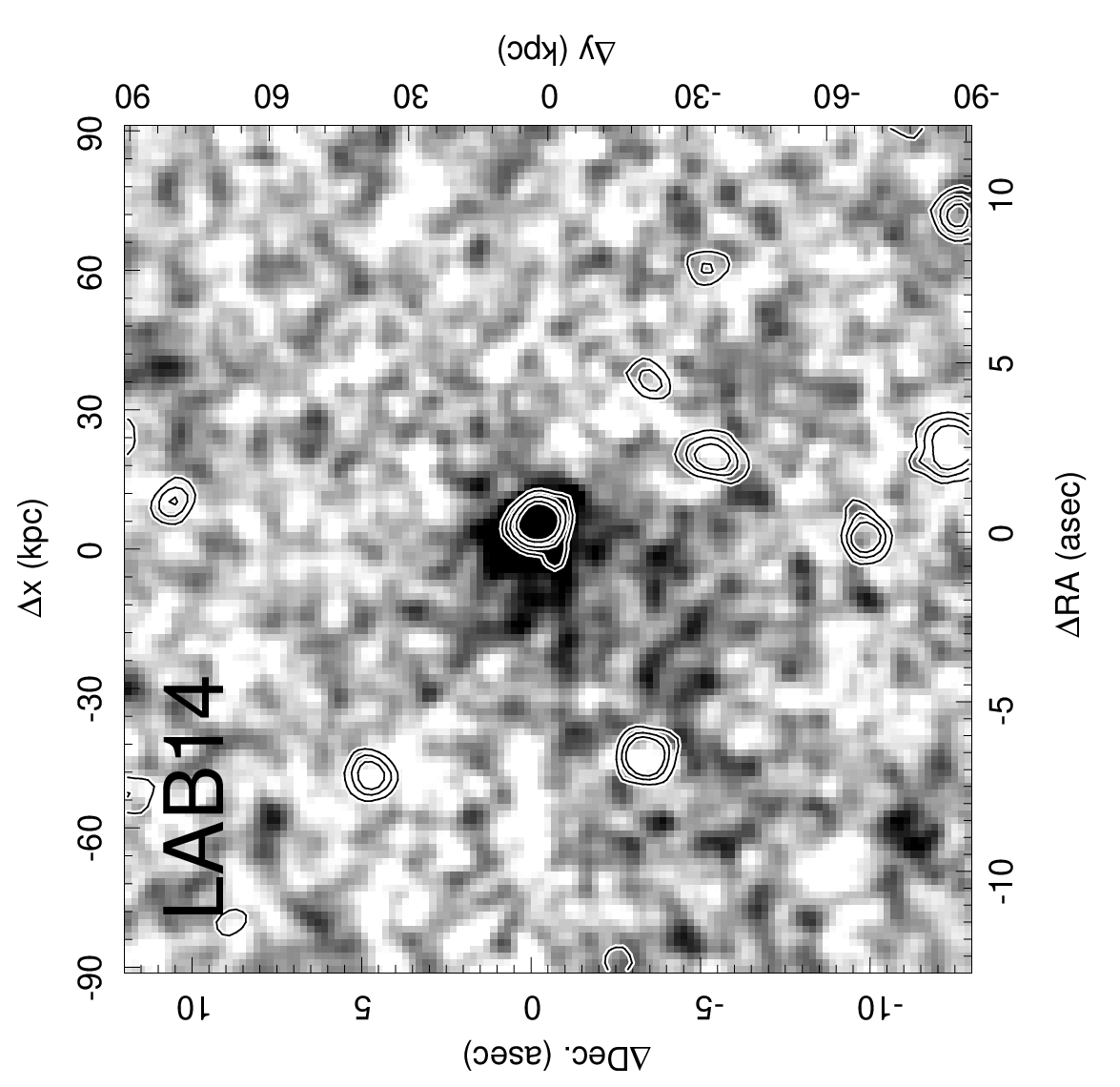}\end{turn}
\begin{turn}{-90}\includegraphics[scale=0.2]{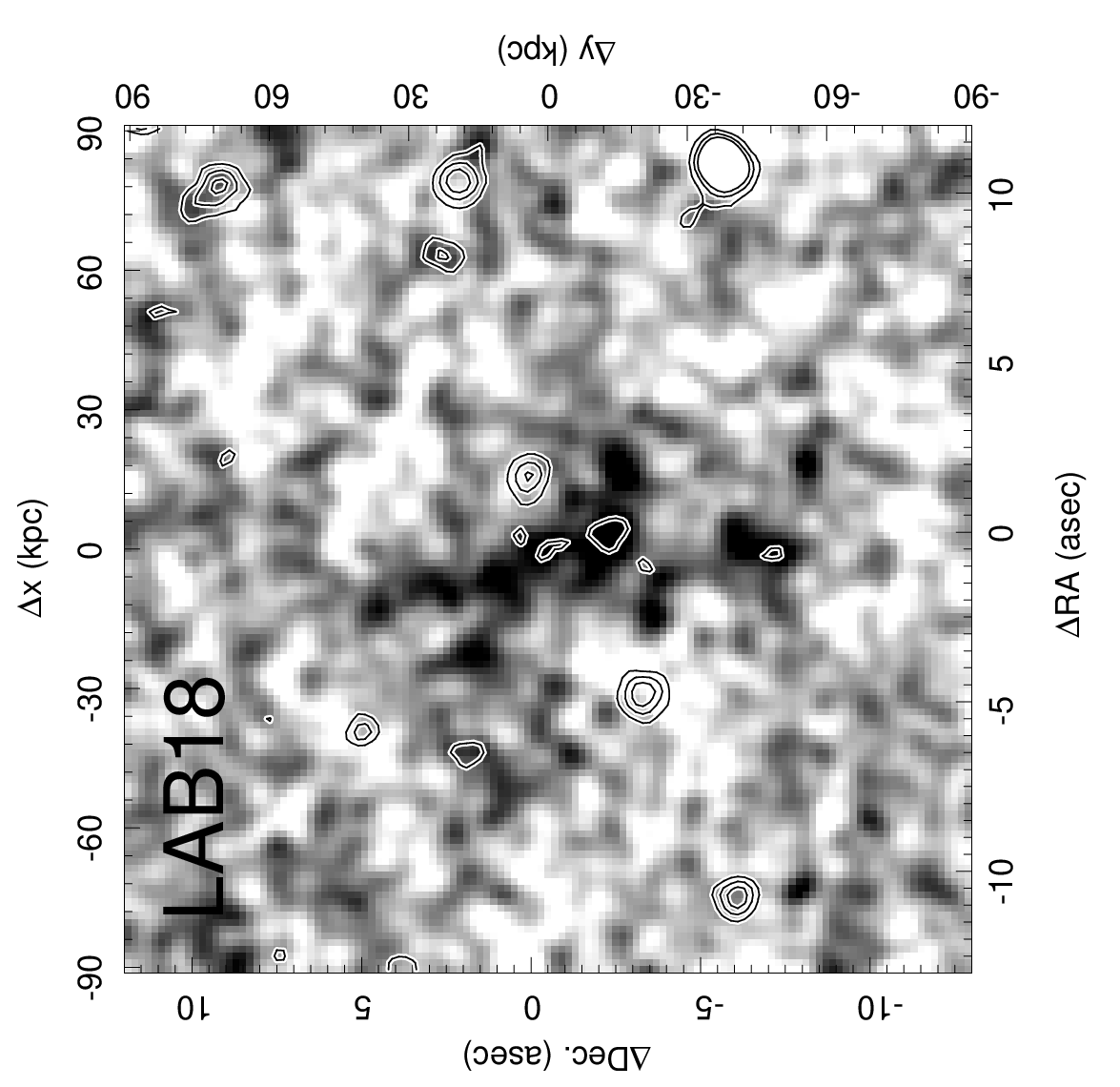}\end{turn}

\bigskip
\bigskip

\begin{turn}{-90}\includegraphics[scale=0.2]{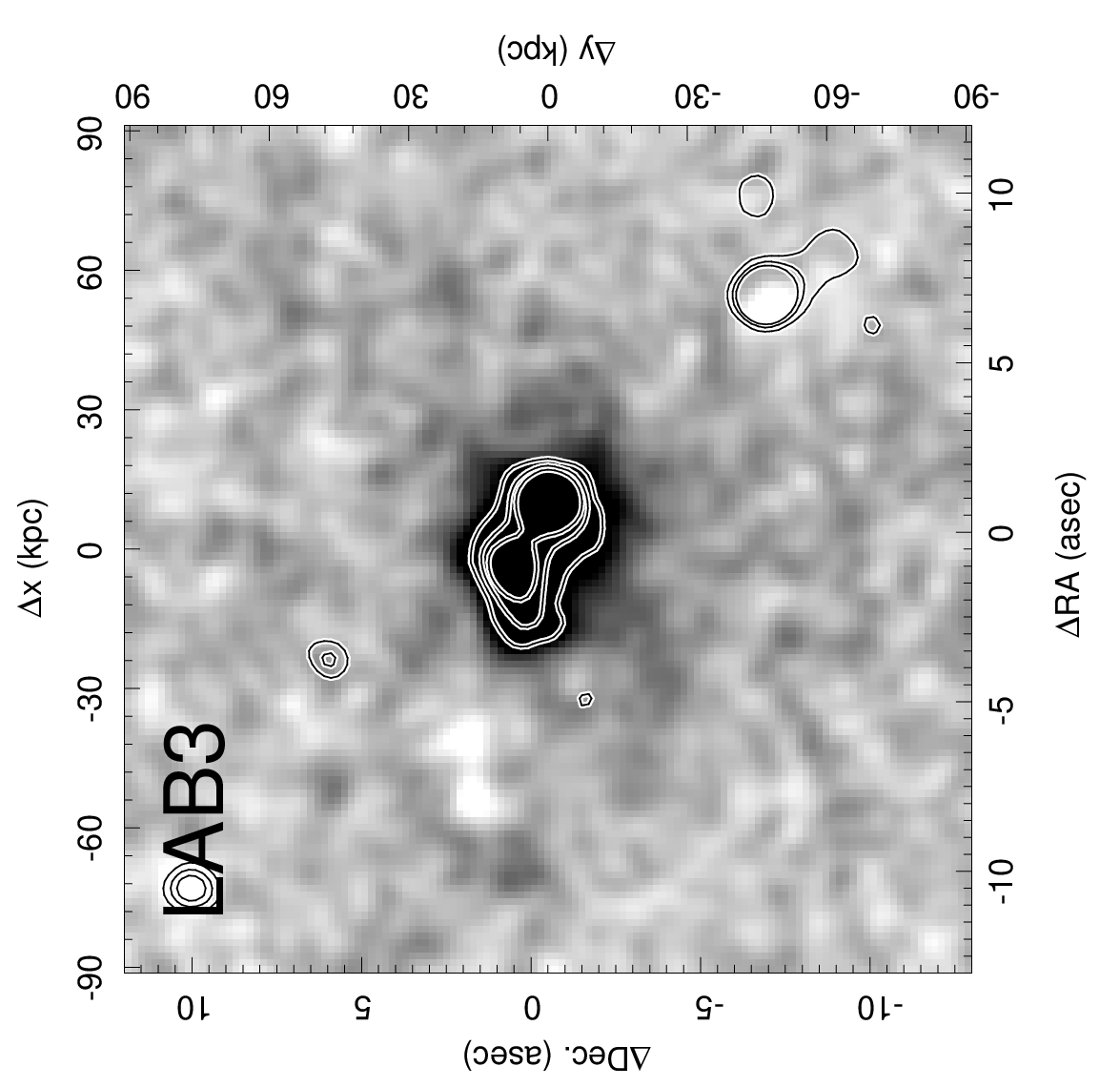}\end{turn}
\begin{turn}{-90}\includegraphics[scale=0.2]{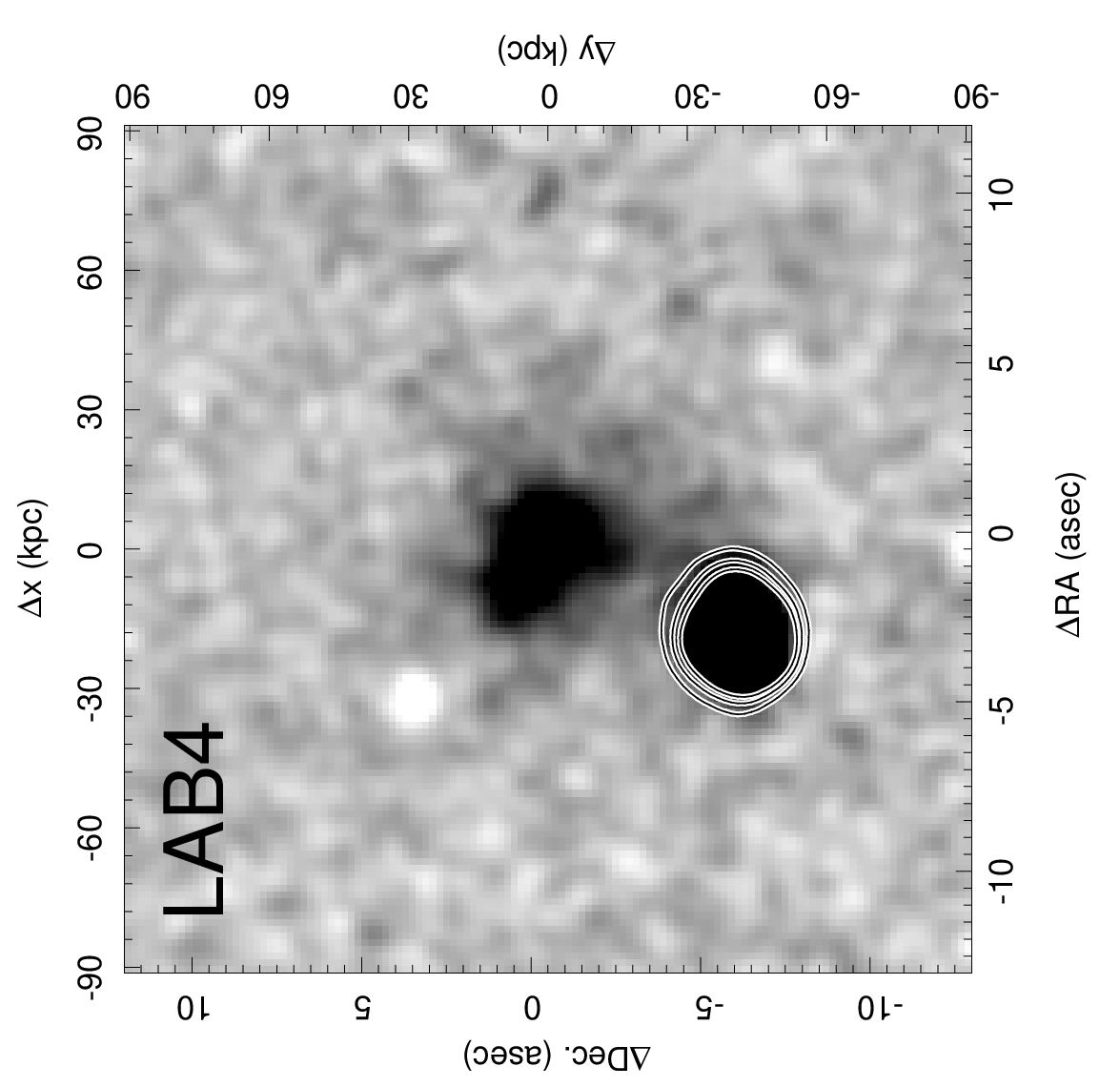}\end{turn}
\begin{turn}{-90}\includegraphics[scale=0.2]{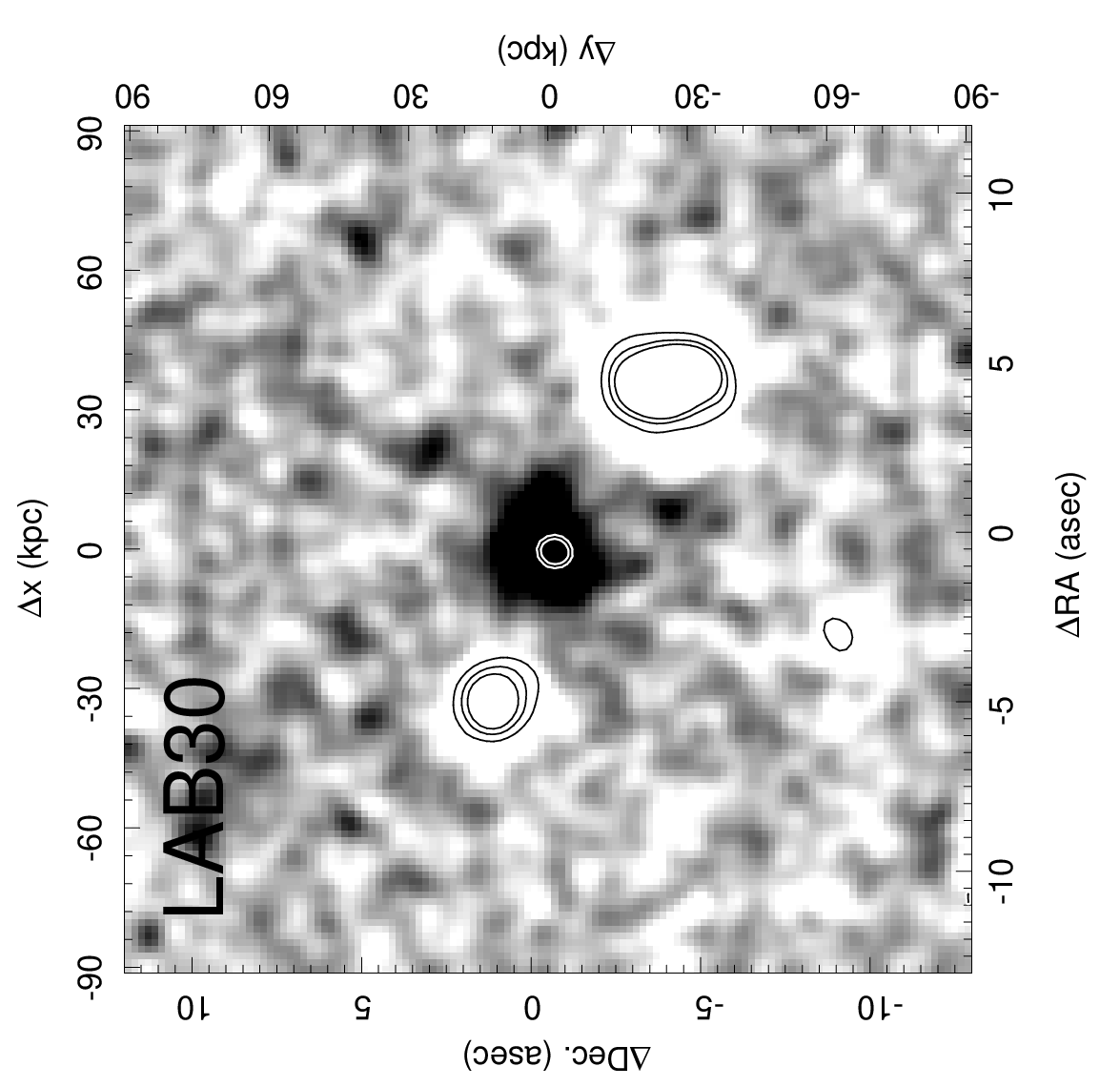}\end{turn}
\begin{turn}{-90}\includegraphics[scale=0.2]{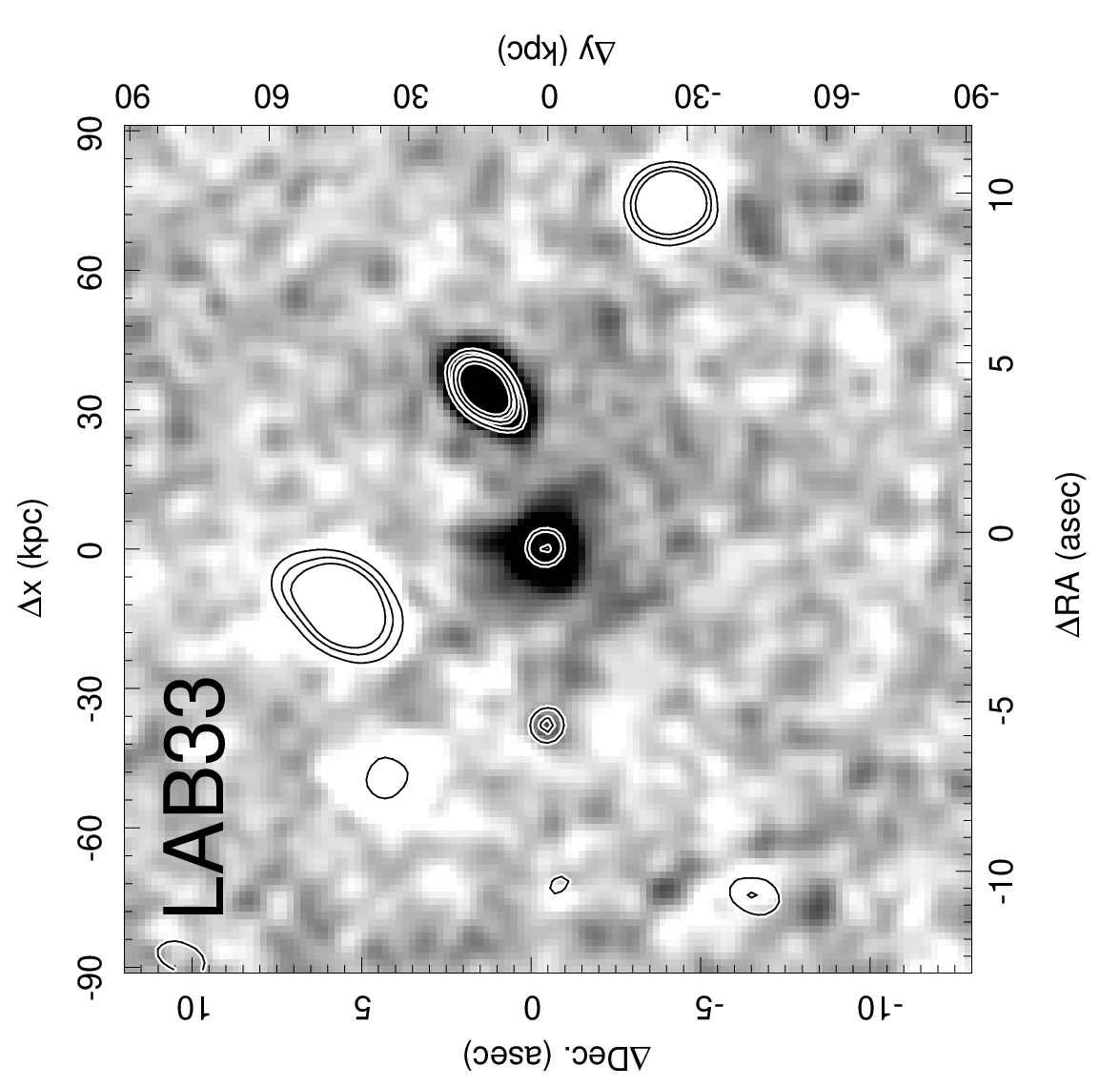}\end{turn}
\begin{turn}{-90}\includegraphics[scale=0.2]{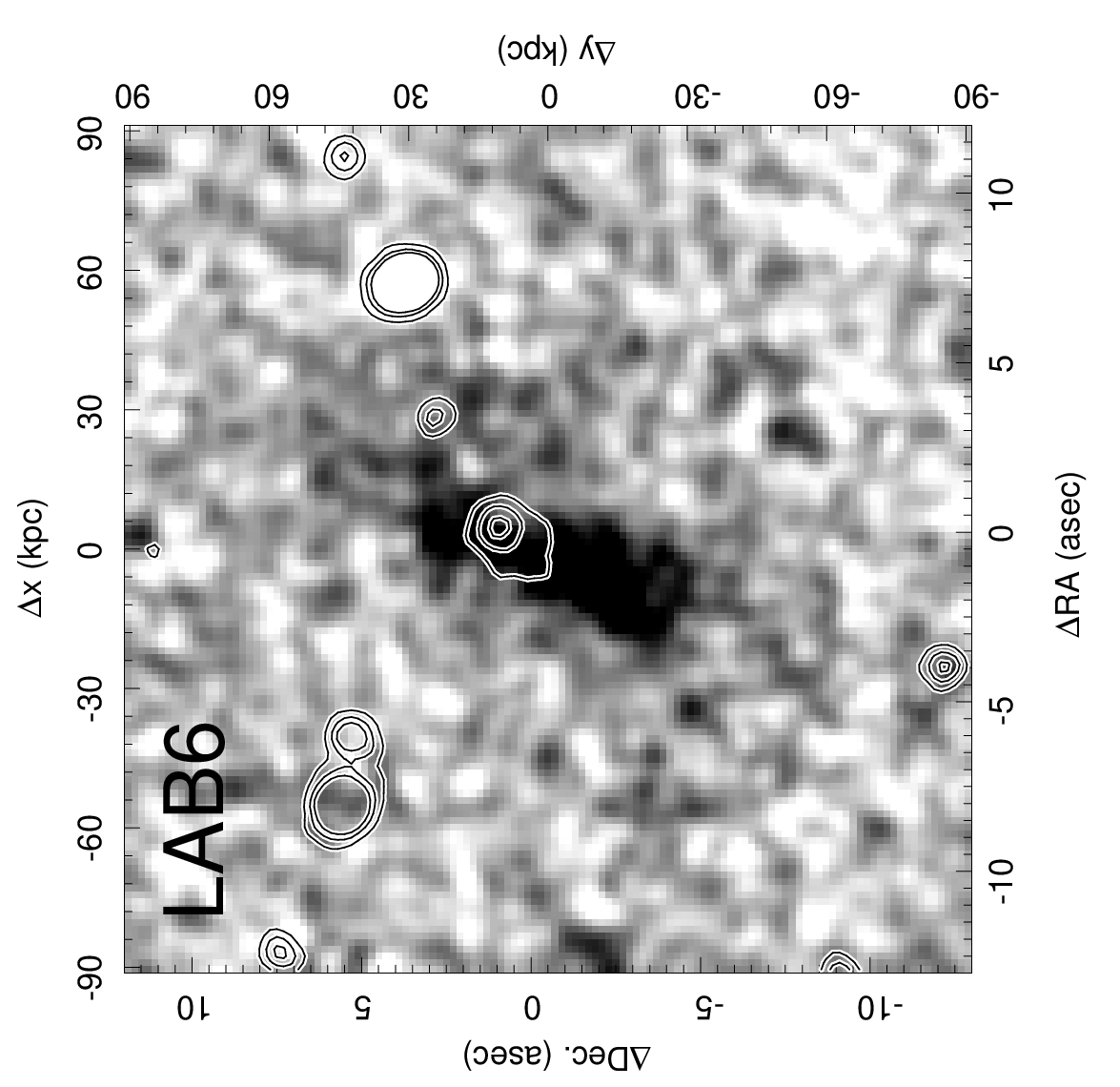}\end{turn}
\begin{turn}{-90}\includegraphics[scale=0.2]{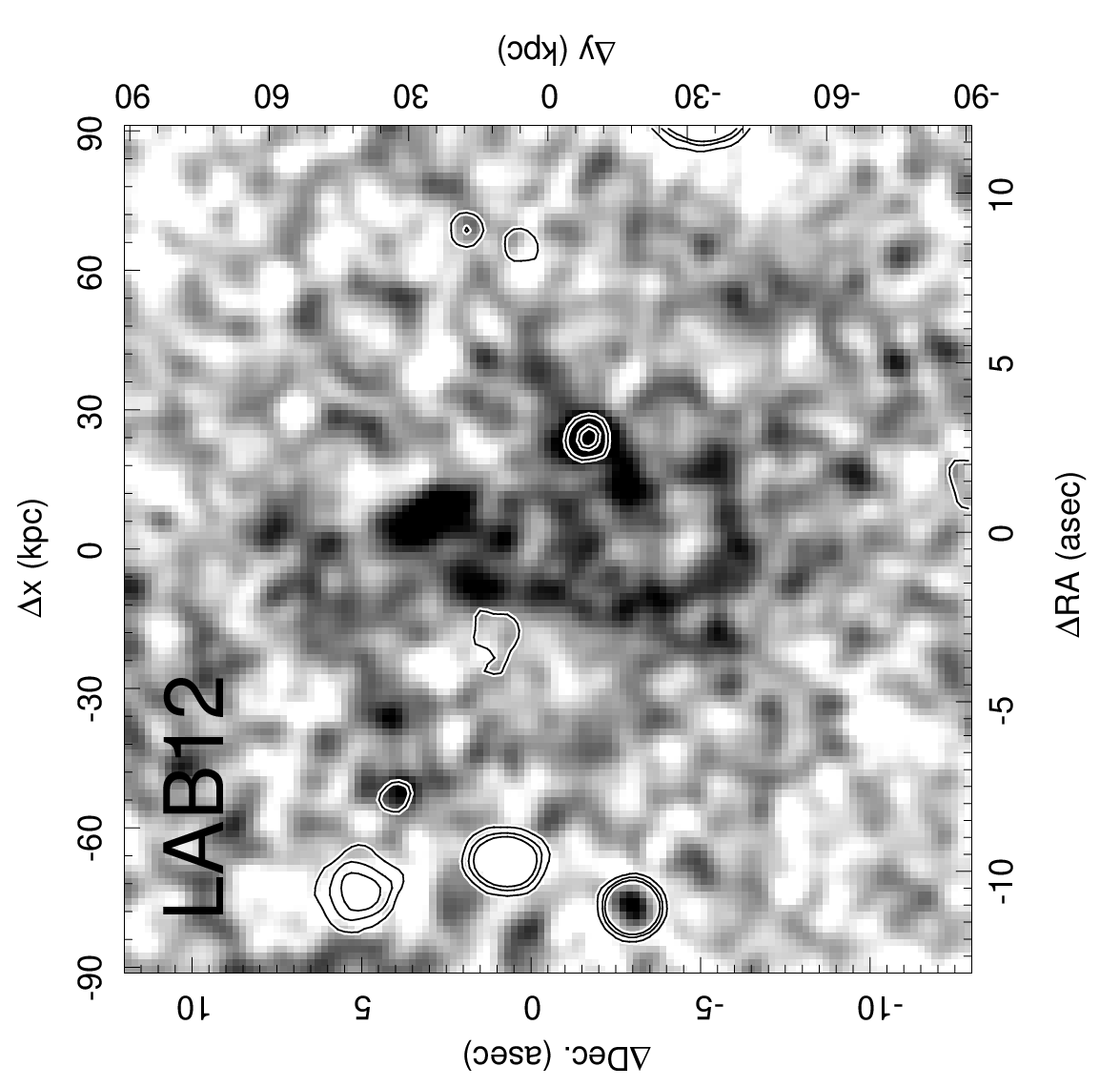}\end{turn}
\begin{turn}{-90}\includegraphics[scale=0.2]{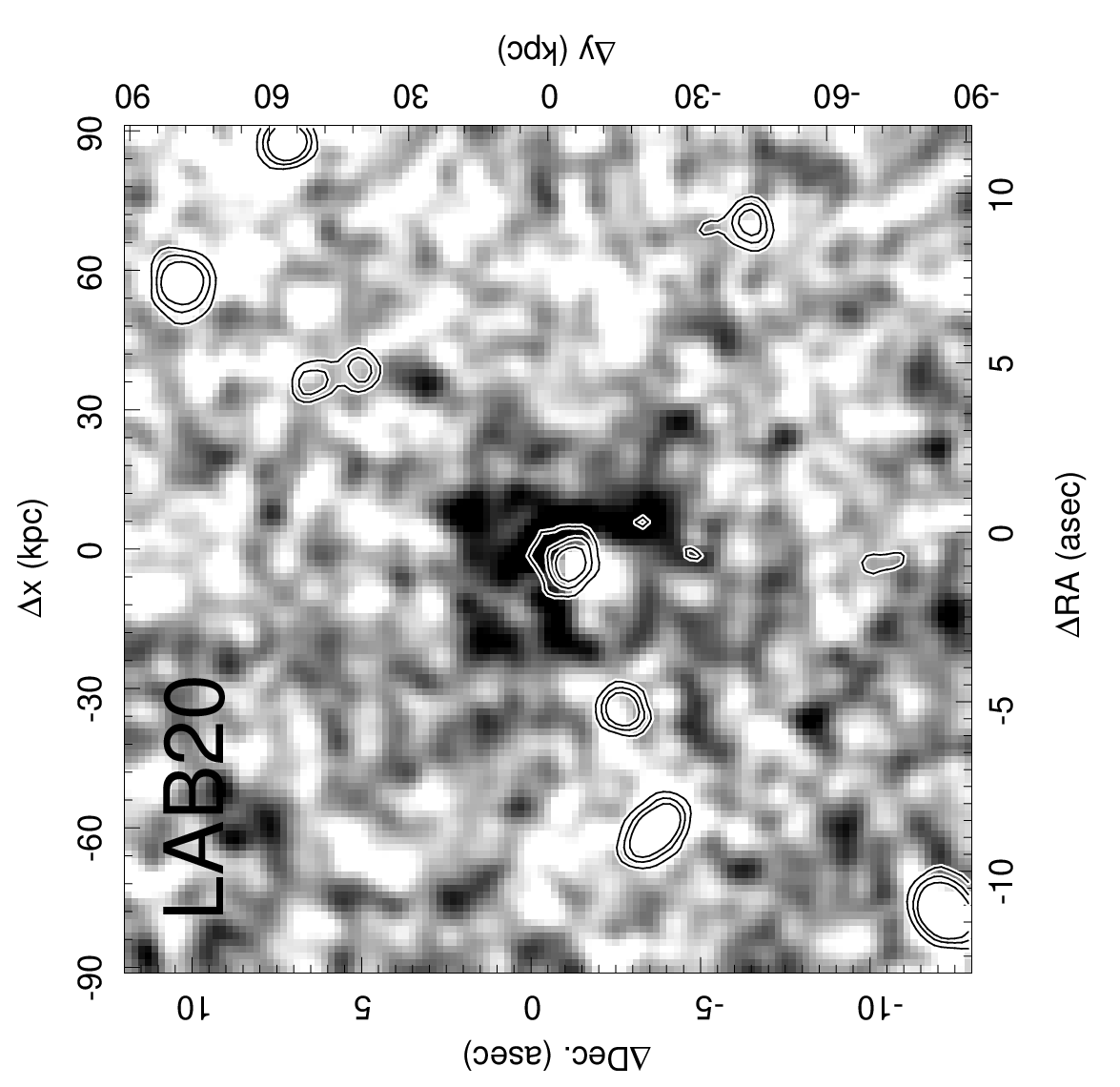}\end{turn}
\begin{turn}{-90}\includegraphics[scale=0.2]{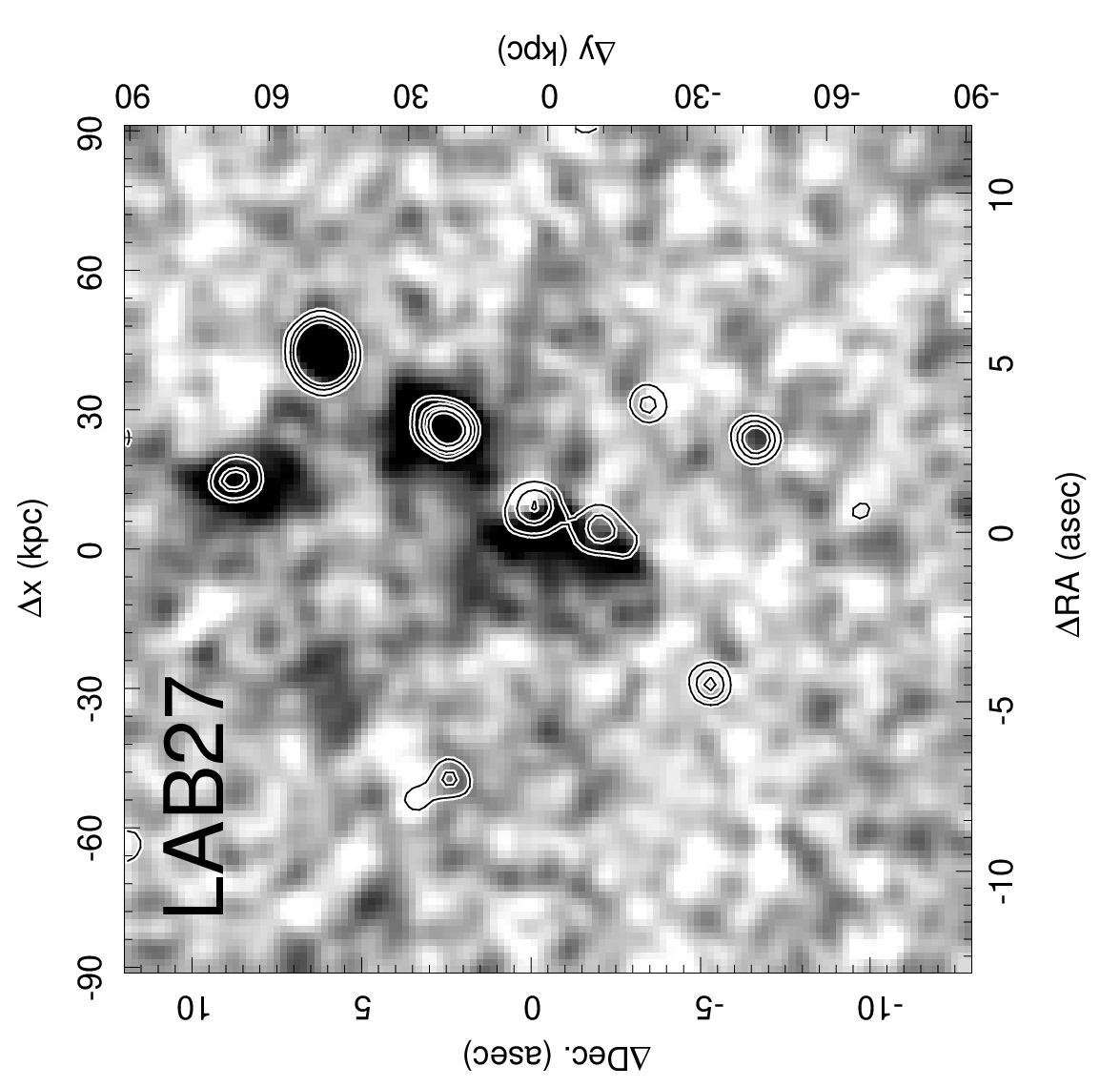}\end{turn}

\caption{Narrow-band Ly$\alpha$ imaging of a subset of the LABs in our submm photometry sample with submm emission (top row) and without detected submm emission (bottom two rows). In this figure we divide the non-detected LABs into two morphological categories: in the top row we show LABs which are compact, with undisturbed haloes; and in the bottom row we show LABs with more complex morphologies -- these objects exhibit more elongated and complex structure than the compact objects. Greyscale denotes the Ly$\alpha$ line emission corrected for continuum emission,
first presented in M04 (see \S\ref{sec:obs} for a description),
while the contours show the location of continuum emission at levels of $\sim3$, 5 and
7$\sigma$. The intensity scales are identical, each panel is
$25''\times25''$ ($\sim190$\,kpc\,$ \times 190$\,kpc) and have North at the top and
East to the left. The panels are centred on the coordinates listed in
Table~1.}
\label{fig:labplots1}
\end{figure*}

To look at this question in more
detail we also study the morphologies of the
Ly$\alpha$ haloes for the detected and non-detected submm sources (Figure~\ref{fig:labplots1}).
Generally the LAB's morphologies fall into two classes -- those with
an elongated or apparently disturbed halo (e.g.\ LAB5, 12, 18), and
those which are compact and less disturbed (e.g.\ LAB3, 10, 30). The
submm-detected LABs have a range of morphology: LAB5 and LAB18 are
elongated and diffuse, whereas LAB10 and LAB14 are more compact and
circular. To show that the morphologies are not a significant factor
for the submm emission we note that each of the four submm-detected
LABs has a morphological counterpart in the non-detected sample
(e.g.\ LAB5 versus LAB6 or LAB10 versus  LAB4)

In summary, we can find no property of the Ly$\alpha$ emission from the
LABs which correlates strongly with their submm detection rate or flux.
However, given our statistical detection of submm emission
in the whole sample, implying starburst/AGN activity in a large
proportion of LABs, it seems likely that this bolometrically-luminous
activity must have some relationship with the LAB haloes. 
For example, if
a wind is responsible for the emission, then we may be
observing various stages of LAB evolution, governed by the nature and
environment of the submm source.

%
%

\subsection{Environmental properties of LABs}
\label{ssec:env}

To determine any trend in the environments of the submm-detected LABs, we
compare them to the local surface density of 283 Ly$\alpha$ emitters
(LAEs) at $z=3.09$
in SA\,22 from M04 in Fig.~\ref{fig:skyplot}. The contours compare the local density of LAEs to the average density over the field-of-view. The
LAE distribution was convolved with a Gaussian with a co-moving size of $\sigma=2.8$\,Mpc. The bold contours indicate regions of greater than average local surface density, clearly showing the filamentary large-scale structure in the proto-cluster.  

Although LAB1 resides close to a node in the large-scale
structure (Matsuda et al. 2005 in prep.), we find no tendency for the
other submm-detected LABs to reside in
higher density local environments than the average LAE, 
suggesting that this behaviour isn't reflected in the wider sample.  
Indeed, if anything it might be argued that our new submm-detected
LABs lie in lower-density regions on average, although this is not
statistically significant. We note that although the underlying density field may not be correlated with the positions of LABs, the local density may have an effect on the properties of LABs, which may account for the range of sizes and luminosities we observe.

\section{Discussion}
\label{sec:dis}

We now discuss the wider insights which our observations
provide into the properties and origin
of LABs and their environments. We concentrate on the interaction of winds generated from luminous activity within the embedded SMGs with the ICM as the most likely source of Ly$\alpha$ emission.

\subsection{LAB formation mechanisms}
\label{ssec:mecha}

%
%
\begin{figure}

\begin{center}
\begin{turn}{-90}
\includegraphics[scale=0.47]{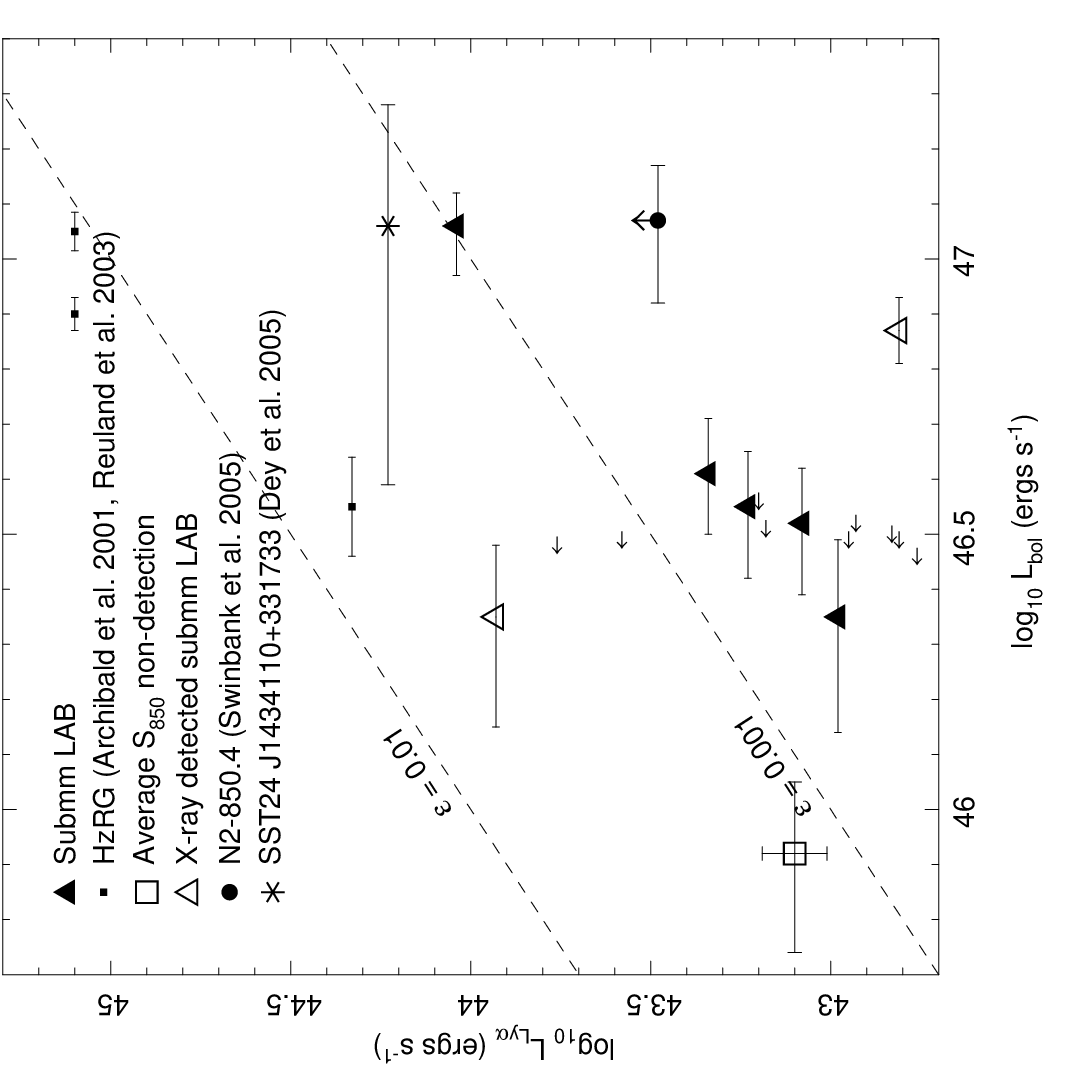}
\end{turn}
\end{center}
\caption{Ly$\alpha$ luminosity versus bolometric luminosity (assuming a
modified black-body with $T_{\rm d} = 40$\,K, $\alpha=4.5$, $\beta=1.7$) for
submm-detected LABs and potentially related objects.  Note that we plot
LAB30 as a submm-detected LAB, and 3$\sigma$ upper
limits are plotted for the remainder of the photometry sample. Also
shown is the flux obtained from stacking the  submm-undetected LABs
from the photometry sample. The starburst galaxy N2-850.4 (Smail et al.\
2003; Swinbank et al.\ 2005) is also shown, along with the {\it Spitzer}
detected source SST24 J1434110+331733: both these objects have been
identified with extended Ly$\alpha$ haloes. The dashed lines show a simple
model where the Ly$\alpha$ luminosity is determined as a fraction
$\epsilon$ of the bolometric luminosity. We show this model for the
scenarios $\epsilon = 0.01$ and $\epsilon = 0.001$.    We also
plot three submm-detected, high-redshift radio galaxies with extended
Ly$\alpha$ emission for comparison.}
\label{fig:bol}
\end{figure}

In Figure~4 we plot the Ly$\alpha$ luminosity against the bolometric
luminosity calculated for the submm-detected LABs -- LAB1, 5, 10 and
14. We have also plotted the flux of the submm-undetected
LABs from our stacking analysis and two objects
which are
proposed to be similar to LABs: the SCUBA galaxy N2-850.4 (Keel et al.\ 1999, Smail et al.\ 2003) and the
{\it Spitzer} identified source SST24 J1434110+331733 (Dey et al.\
2005). We discuss these analogous systems further in \S
\ref{ssec:general}.\footnote{We note that the Extremely Red Object 2142-4420 B1 (Francis et al.\ 2001) also has been identified with an
extended halo ($L_{\rm Ly\alpha}\sim10^{44}$\,ergs\,s$^{-1}$). Those
authors suggest that the embedded galaxy's FIR emission could be of the
order $5\times10^{47}$\,ergs\,s$^{-1}$, although a precise far-infrared
luminosity is not available. If this predicted FIR luminosity is
correct
then it would
place the source at a similar location to SST24 in this Fig.~\ref{fig:bol}.} 
The bolometric luminosities of the LABs are calculated by
integrating over a modified black-body curve with a dust temperature of
40\,K and $(\alpha,\beta)$ = $(4.5,1.7)$. 

This analysis allows us to
examine any causal connection between the bolometric output of the LAB, and the
luminosity of the Ly$\alpha$ halo. As Fig.~4 shows there is  a
weak trend between $L_{\rm bol}$ and the luminosity of the
extended Ly$\alpha$ halos for the sample -- in the
sense that the most bolometrically-luminous
sources have more luminous associated Ly$\alpha$ halos.
 
We adopt a simple model in which the Ly$\alpha$ emission traces a
fraction $\epsilon$ of the bolometric luminosity, where $\epsilon$
combines the fraction of energy output by star formation or AGN which
generates a wind and the fraction of that wind energy which produces
Ly$\alpha$ emission: $L_{\rm Ly\alpha} = \epsilon L_{\rm bol}$ In
Figure~4 we plot this behaviour for values of $\epsilon$ corresponding
to 1 and 0.1 per cent of the bolometric output. A fractional
output around $\sim 0.1$ per cent
best describes the LABs, indicating that only a small proportion of the
bolometric output from the obscured activity in these
galaxies is required to support these
luminous haloes.   The weak trend we see is consistent with a
direct causal link between the extended Ly$\alpha$ emission and
that from the obscured sources detected in the submm.  The most likely
mechanism which could provide this link is 
winds (Ohyama et al.\ 2003) and we suggest that winds are the major
cause of the extended Ly$\alpha$ emission seen in the LAB population.
This proposal would indicate that  LABs may be found around a large range of
the submm population and we explore this idea further in
\S\ref{ssec:general}.

Clearly the trend shown in Fig.~4 is not a tight correlation.
There are several possible explanations for this -- including
different efficiencies for coupling energy to the halo from
different sources (e.g.\ AGN or starbursts), potentially 
cyclic
activity, environmental influences and observational uncertainties. 
We provide a more quantitative discussion 
the mechanics of emergent winds and jets, and possible environmental
influences on  luminosity of Ly$\alpha$ halos in the next section. We note that the two LABs with possible X-ray counterparts -- LAB2 and 18
-- are not well described by the general trend. These are the two LABs
whose tentative X-ray detections suggest they may contain obscured AGN (upper limits for the remainder of the population are consistent with the measured fluxes and luminosities of these two marginal detections).
This might explain why they depart from the trend in Fig.~\ref{fig:bol}
(i.e.\ AGN may interact differently with the ICM than starbursts).
However, the fact that they lie respectively above and below the trend
suggests that other factors may be at work, in particular their local
environments. Fig.~\ref{fig:skyplot} shows that LAB18 lies in one of the
lowest density regions in our survey, while LAB2 is close to the peak in
the LAE distribution. LAB18 has an submm flux $\sim 3\times$ that of LAB2,
but LAB2's Ly$\alpha$ luminosity is over an order of magnitude greater
than LAB18.  This suggests that although a powerful embedded submm source
might be required for the generation of an LAB, it may be the local
environment (i.e.\ the density of ambient intracluster gas) which governs
the efficiency of coupling of the bolometric output to the luminosity of
the halo. This is a natural explanation for the apparent diversity of
these objects' properties. Unfortunately, our survey is too sparse to
conclusively test for environmental influences on the properties of
submm-bright LABs. 

To demonstrate the potential variation in L$_{\rm Ly\alpha}$--L$_{\rm bol}$
which may arise from AGN-driven activity we also 
plot on Fig.~4  three HzRGs from Reuland et al.\ (2003)
with detected 850\,$\mu$m emission  from Archibald et al.\ (2001),
converted to our cosmology.   These galaxies show a much higher
ratio of Ly$\alpha$ emission at a fixed bolometric output,
compared to the LABs indicating the  role of powerful
radio jets and possible 
contributions from inverse-Compton ionization (Scharf et al.\ 2004) in
addition to the starburst or AGN activity, shown by the LABs.

\subsection{Origin of the wind}

In this paper we have used the term ``wind'' to refer to the
interaction between any outflowing material 
whatever its origin and the intergalactic medium
(or the young, intracluster
medium in the case of SA\,22). 
Are there any observational signatures to distinguish the feedback of
starbursts and AGN on the ICM? AGN are very localised -- they can
impart energy into the galaxy via direct irradiation or jets: highly
collimated outflows from matter accretion onto a black hole. Starbursts
are more complex, in the sense that the energy release can be
distributed over a wider volume (in giant molecular clouds), and
comprises of the output of young massive stars, and a contribution from
an enhanced SNe rate.
For an AGN, on larger scales the
most likely mechanism for generating an outflow is expected to be jets
breaking out of the galaxy and into the IGM. These outflows entrain
material from the galaxy (in the case of SMGs this would include dust)
and the IGM, and create a shock front when they become supersonic in
the ambient medium. It is this shock which can heat the ambient gas to
ionising temperatures, and therefore dissipate the jet's energy. In
this situation the observations will be orientation specific, such that
a face-on jet induced LAB will be compact and bright, whereas a side-on
jet might resemble an elongated structure. The morphological diversity
of the LABs could be explained by this variety of mechanisms, with both
these observational classes being found in our sample (e.g.\ LAB10 and
LAB18 respectively), although it is unclear whether an AGN, starburst,
or combination of the two is responsible. We note that without detailed
spectroscopic evidence to unravel the kinematics of these systems,
caution must be exercised in our wind analysis: similar structures
could be easily explained by the distribution of tidal material
resulting from a merger.

Starburst-driven winds can emerge from a galaxy and interact with the
IGM in a similar way to an AGN jet. Although not as concentrated, these
winds can still be collimated, and resemble jets with a very wide
opening angle (at least for local far-infrared luminous galaxies, e.g.\
Heckman et al.\ 1990). For these collimated winds, a similar
orientation dependence governs observations, but the loose collimation
allows the formation of the extended diffuse haloes present in several
of these LABs (e.g.\ LAB5, LAB6). The escape of galactic winds from
dusty, high-redshift starburst galaxies might be different to their
lower-redshift counterparts (e.g.\ M82), and we must take that into
account when considering these scenarios. Given that it is possible for
AGN and starbursts to generate similar morphological structures in
LABs and so with our present observational tools we cannot clearly
distinguish between AGN- and starburst-powered winds in the LABs.
However, if the obscured activity seen in the LAB population is
exactly analogous to that in typical submm galaxies at this epoch,
then from detailed studies of the AGN in the latter population
we can conclude that it is likely that star formation is
responsible for the winds in the LABs (Swinbank
et al.\ 2004; Alexander et al.\ 2005).

\subsection{Energy injection, mass loss and age}
\label{ssec:ei}

Heckman et al.\ (1990) discuss the mechanics of emergent superwinds from
the discs of a sample of low-$z$ far-infrared galaxies. Those authors
present a simple model for the energy injection rate (into some ambient
medium) from the inflation of a bubble due to the combined effect of
stellar winds from massive stars and the detonation of supernovae:
$dE/dt \sim
3\times10^{41}r^2v^3n_0\phantom{x}\rm{ergs}\phantom{x}\rm{s^{-1}}$.
Where $r$ is the radius in kpc of the bubble, $v$ is the wind velocity
in units of 100\,km\,s$^{-1}$ and $n_0$ is the density of the
undisturbed medium just outside the bubble in cm$^{-3}$. Applying this
to our sample; if we take the radius of the bubble in LABs to be the
extent of the emission (in this case between $\sim$10--100\,kpc with a
median of 22\,kpc), a wind velocity of $\sim1000$\,km\,s$^{-1}$
(characteristic of the escape velocity from a massive galaxy) and a
density of 1\,cm$^{-3}$ (Shull \& McKee 1979) then the energy injection
rate is of the order $10^{45}$\,ergs\,s$^{-1}$. The typical
luminosities of LABs in this sample are approximately two orders of
magnitude lower than this, suggesting that the conversion of bolometric
energy to Ly$\alpha$ emission (via winds) is an inefficient
process.  The conversion efficiency from bolometric to
Ly$\alpha$ luminosity derived
from the trend in Fig.~\ref{fig:bol} is $\sim0.1$ per cent, which would support this
interpretation. We note that the effect of the density of the ambient medium
on the injection rate is vital in this analysis -- a lower ambient
density favours a lower coupling of bolometric luminosity (that
generated by the starburst or AGN) and a wind into the IGM. This would
provide a natural explanation of 
the fact that there is significant scatter in the trend between $L_{\rm
bol}$ and $L_{\rm Ly\alpha}$.  For example LAB18, which resides in a
much lower density environment to the rest of the LABs has a high
bolometric luminosity (second only to LAB1), but a much lower
Ly$\alpha$ luminosity. 

This simple wind model allows us to estimate the age of the starburst
(or AGN activity) by comparing the extent of the haloes with the
velocity of the superwind. For the figures estimates above, these LABs
could have starbursts of the age $\sim 20$\,Myr -- although could be up to
$\sim 100$\,Myr for the largest LABs. Caution must be exercised for
this interpretation due to the fact that a halo might remain as an
emission line nebula for some time after luminous activity has ceased,
or when cyclic models of activity are considered, a succession of which
may have illuminated a halo at some former time, but which is now
fading. Nonetheless, starbursts of this age are consistent with
estimates for other SMGs (e.g.\ $>10$\,Myr for N2-850.4, Smail et al.\
2003).

\subsection{LABs around the submm population in general?}
\label{ssec:general}

About 20 per cent of the LABs in our (photometry) sample contain
strong submm sources.  Combined with the apparent trend between
Ly$\alpha$ and FIR emission, and the morphological analysis, this
supports the interpretation that a galactic superwind is responsible
for the Ly$\alpha$ emission. We speculate that giant Ly$\alpha$
emission-line haloes may be a feature of the general submm population.

Spectroscopic observations of SCUBA galaxies based on
radio-identification (Chapman et al.\ 2005) have yielded impressive
results, with the redshifts of $\sim$100 galaxies currently known. An
interesting feature of the spectra is the presence of strong Ly$\alpha$
emission lines which aids in the measurement of redshifts. For example,
in Chapman et al's spectroscopic survey of SMGs, of the 73 successful
identifications, 34 per cent of them were made primarily with the
Ly$\alpha$ line. This runs counter to the expectation that the active
regions in submm galaxies should be heavily obscured by dust resulting
from an intense starburst, and hence these regions are unlikely to be
strong Ly$\alpha$ sources since dust efficiently destroys Ly$\alpha$
photons. The presence of the Ly$\alpha$ line then suggests two
possibilities: (a) the dust coverage is patchy, and Ly$\alpha$ photons
generated within the galaxy escape through holes in the distribution
(Chapman et al.\ 2004); or (b) the Ly$\alpha$ emission originates well
away from the dust in a halo coupled to the galaxy. This latter
scenario is consistent with the LAB picture and would suggest that
extended Ly$\alpha$ haloes should be found around many SMGs. 

There already exists some evidence for extended Ly$\alpha$ halos
around luminous, far-infrared galaxies at high redshifts: the first submm-selected galaxy, SMM J02399-0136 (Ivison et al. 1998), also has a large Ly$\alpha$ halo, covering virtually all of the 15$^{''}$ slit used by those authors. Similarly, Smail et al.\ (2003) detect four SMGs within
$\sim 1$\,Mpc of the radio galaxy 53W002 ($z=2.39$), and identify one of them,
SMM J17142+5016 as a narrow-line AGN with an extended Ly$\alpha$
halo. Finally, the SCUBA galaxy N2-850.4 (SMMJ163650.43+405734.5 Scott et al.\ 2002; Ivison et al.\ 2002; Smail et al.\ 2003) is a hyperluminous class IR
galaxy undergoing a starburst at $z=2.385$. There appears to be a diffuse H$\alpha$ 
(and perhaps Ly$\alpha$) halo surrounding the galaxy,
which although apparently compact, may extend to large distances with a
surface brightness below the limit of the observations. For comparison, we plot N2-850.4 in Figure~\ref{fig:bol}, using a lower
limit for the Ly$\alpha$ luminosity of $>3\times 10^{43}$
ergs\,s$^{-1}$ (Swinbank et al.\ 2005), and $L_{\rm bol}\simeq L_{\rm
FIR} = 3.1\times 10^{13} L_{\odot}$ (Chapman et al.\ 2005).  

Very recently, 
Dey et al.\ (2005) identified a {\it Spitzer Space Telescope} source,
SST24 J1434110+331733, which is surrounded by a 200\,kpc-diameter
Ly$\alpha$ nebula at $z\sim 2.7$.  The nebula has $L_{\rm
Ly\alpha}\sim1.7\times10^{44}$ ergs s$^{-1}$ and the nebula hosts a
bright mid-infrared source ($S_{\rm 24\mu m}$ = 0.86 mJy), with an
implied FIR luminosity of $L_{\rm FIR}\sim 1$--$5\times 10^{13}$
$L_\odot$, although the estimate is uncertain as it involves an
order-of-magnitude extrapolation.  The size and luminosity of the
Ly$\alpha$ halo, combined with the potentially hyperluminous source
lying within it make this a close analog of LAB1. We plot it for
comparison in Fig.~\ref{fig:bol}, and note that it follows the trend shown by the
majority of submm-detected LABs.  This close association between a
luminous, obscured source and a highly-extended Ly$\alpha$ halo further
supports our proposal that LABs may be a common feature of the high-redshift far-infrared population.

For completeness, however,
we should note that
our survey of SA\,22 contains at least one submm source which does not
appear to be associated with an extended halo. SMMJ221735.15+001537.2
is detected with $S_{850} = 6.3\pm1.3$ mJy (Chapman et al.\ 2005) at $z
= 3.098$. Ly$\alpha$ is detected for this system, but it is not
sufficiently extended to qualify as a LAB. The lack of spatial extent
would be consistent with the picture where this is a young submm source
and the wind has only just started to propagate into the IGM, thus
resulting in a small or faint halo. Alternatively, as discussed above, local environmental conditions could play a role in the lack of emission we observe.

\subsection{Star formation in the SA\,22 region}
\label{ssec:ssa22}

Our detection of submm emission from the average
LAB in our survey within the SA\,22 field also allows us to compare the star formation
rate density (SFRD) within this structure to that of the field at this epoch. To estimate the
volume containing the $z=3.09$ structure in  SA\,22 we assume that the star-formation is contained 
within a co-moving sphere of apparent size 15$^{'}$, corresponding to a volume of $\sim$1.1$\times10^4$\,Mpc$^3$.

Directly integrating the submm emission from the six detected LABs (1,
2, 5, 10, 14, 18) and the other SMG known to lie at $z=3.09$ in this
field (Chapman et al.\ 2005) we derive a total SFR in submm galaxies
of $\sim 1.4\times10^{4}$\,$M_\odot$\,yr$^{-1}$.
This yields a star formation rate
density (SFRD) in SA\,22: $\rho_{\rm  SFR}>1.3$\,$M_\odot$\,yr$^{-1}$\,Mpc$^{-3}$. Alternatively, given our average submm flux for 
LABs of $\sim$3.1\,mJy, the entire M04 catalog of LABs would yield a
SFRD of $\rho_{\rm SFR} > 3.2$\,$M_\odot$\,yr$^{-1}$\,Mpc$^{-3}$ (star
formation rates are calculated assuming $L_{\rm bol} \sim L_{\rm IR}$ after
Kennicutt 1998). 

A lower limit to the average SFRD at $z\sim 3$ is provided by
UV-selected
surveys of LBGs at this epoch, giving a SFRD of
$\geq 0.1$\,$M_\odot$\,yr$^{-1}$\,Mpc$^{-3}$
(Steidel et al.\ 1999) after accounting for dust extinction and extrapolating the
LBG luminosity function to faint limits.  Alternatively, we can use
submm surveys to attempt to determine the SFRD from a
bolometrically-selected
sample at this epoch, although at the cost of extrapolating the 
submm luminosity function.  The spectroscopic survey of SMGs by
Chapman et al.\ (2005) indicates an average SFRD at $z\sim 3$ 
of $\sim 0.8$\,$M_\odot$\,yr$^{-1}$\,Mpc$^{-3}$.  The estimates of
the SFRD within the SA\,22 structure indicate significant acceleration of early
star-formation
in this overdense region (Ivison et al.\ 2000; Stevens et al.\ 2003;
Smail et al.\ 2003).

\section{Conclusions}
\label{sec:con}

We have presented results from a submm SCUBA survey of a sample of 23
giant, Ly$\alpha$ emission line nebulae -- Ly$\alpha$ Blobs (LABs) --
at $z=3.09$ in the SA\,22 protocluster. In addition to the previously studied LAB1
and LAB2 which have been shown to contain submm sources, we identify a
published submmm source with LAB14 and from our photometry
observations detect a further three (LAB5, 10, 18) with 850\,$\mu$m
fluxes $>3.5\sigma$. We conclude the following:

\begin{itemize}

\item We have tripled the number of LABs with detected submm emission
in the SA\,22 overdensity to a total of six sources.
The $z = 3.1$ protocluster in SA\,22 is thus the
richest association of submm galaxies known.

\item The majority of LABs contain sources which are undergoing an episode of extreme luminous activity; most likely caused by a starburst (although our current limits cannot rule out a contribution from an AGN) and are heavily obscured by dust.

\item The average 850\,$\mu$m flux for the full sample of LABs is
$3.0\pm0.9$\,mJy, corresponding to  star formation rates of the order
10$^3$\,$M_\odot$\,yr$^{-1}$.  
We also find a significant detection of submm emission from those
LABs which are individually undetected in our submm observations.

\item
We estimate that the star formation rate density in
SA\,22 is $\gtrsim$3\,$M_\odot$\,yr$^{-1}$\,Mpc$^{-3}$.  This compares
to recent estimates of 0.8\,$M_\odot$\,yr$^{-1}$\,Mpc$^{-3}$ for the
obscured star formation rate at this epoch (Chapman et al.\ 2003,
2005), indicating an acceleration of star-formation in the SA\,22 structure at this 
early time.

\item We find a trend between the Ly$\alpha$ luminosity of the haloes
and the far-infrared luminosity of the embedded SMGs. This trend can be simply
modelled in a scheme where a fraction $\epsilon$ of the bolometric
output is converted to Ly$\alpha$ emission. For LABs, this fraction
appears to be of the order $\sim 0.1$ per cent. The existence of
this trend suggests a causal link between the submm activity we
detect and the extended Ly$\alpha$ halos.

\item Galactic-scale `superwinds' generated from the combined effect of
stellar winds from massive stars and the detonation of supernovae in
the starburst, provide a natural explanation of the
properties we see. These mechanisms would allow the intergalactic medium in
the densest regions to be heated and enriched with metals at early times, in
accordance with the observed lack of evolution in intracluster
metallicity in the high density cores of clusters out to $z\sim1$
(e.g.\ Tozzi et al.\ 2003).

\end{itemize} 

Finally we examine the implications of the discovery of a large number
of submm sources associated with LABs and propose that large emission
line haloes might be a common feature of the submm population in general (and
perhaps all active galaxies in rich environments), implying
strong feedback and outflows into the local environment from these
galaxies. Ly$\alpha$ haloes such as these are then excellent candidates
for further studies of feedback systems at high-redshift, and an
essential stage of galaxy evolution.

\section*{Acknowledgements}
The authors wish to thank Mark Swinbank, Richard Wilman, Richard Bower, Cedric Lacey and Masahiro Nagashima for helpful
discussions, and Alastair Edge and Marek Gierlinski for help in analysing the {\it XMM-Newton} data. We thank an anonymous referee for useful comments which improved the clarity of this work. JEG is supported by a PPARC studentship. IRS
acknowledges support from the Royal Society.
This work made use of the Nasa Extragalactic Database
(NED), which is operated by the Jet Propulsion Laboratory, Caltech,
under contract with the National Aeronautics and Space
Administration.


\label{lastpage}

\end{document}